\documentclass[11pt,a4paper]{article}
\usepackage{jheppub}

\usepackage{verbatim}

\usepackage{graphicx} 
\usepackage{amsfonts,amsmath}
\usepackage{hyperref}
\usepackage{tikz, float}
\usepackage{braket}
\usepackage{subcaption}
\usepackage{import}
\usepackage{tikz}
\usepackage{tikz-3dplot}
\usetikzlibrary{shapes,decorations}
\usepackage{pgfplots}
\usepgfplotslibrary{fillbetween}
\usetikzlibrary{arrows.meta}
\pgfplotsset{
    compat=newest
}
\usetikzlibrary{shapes.geometric}
\usetikzlibrary{shapes.misc}
\tikzset{cross/.style={cross out, draw=black,thick, fill=none, minimum size=2*(#1-\pgflinewidth), inner sep=0pt, outer sep=0pt}, cross/.default={3pt}}

\usetikzlibrary{math}
\usetikzlibrary{calc}

\newcommand{\be}{\begin{equation}}
\newcommand{\ee}{\end{equation}}
\newcommand{\ba}{\begin{eqnarray}}
\newcommand{\ea}{\end{eqnarray}}

\newcommand{\JDL}[1]{{\color{green}#1}}

\newcommand{\lb}{\left(}
\newcommand{\rb}{\right)}

\newcommand{\intl}{\int\limits}
\newcommand{\minl}{\min\limits}

\DeclareMathOperator{\arctanh}{arctanh}

\title{Two Splits, Three Ways \\
\large Advances in Double Splitting Quenches}

\author[a]{Joseph Dominicus Lap}
\author[a,b]{Berndt M\"{u}ller}
\author[c]{Andreas Sch\"{a}fer}
\author[c]{Clemens Seidl}

\affiliation[a]{Yale University}
\affiliation[b]{Duke University}
\affiliation[c]{Universit\"{a}t Regensburg}

\emailAdd{Joseph.Dominicus.Lap@yale.edu}
\emailAdd{Berndt.Mueller@duke.edu}
\emailAdd{Andreas.Schaefer@physik.uni-regensburg.de}
\emailAdd{Clemens.Seidl@physik.uni-regensburg.de}

\abstract{In this work we introduce a method for calculating holographic duals of BCFTs with more than two boundaries. We apply it to calculating the dynamics of entanglement entropy in a 1+1d CFT that is instantaneously split into multiple segments and calculate the entanglement entropy as a function of time for the case of two splits, showing that our approach reproduces earlier results for the double split case. Our manuscript lays the groundwork for future calculations of the entanglement entropy for more than two splits and systems at nonzero temperature.}

\begin{document}

\maketitle

\section{Introduction}

\subsection{Motivation}

The disintegration of a strongly interacting quantum system that is overall in a pure quantum state into many pieces that no longer interact with each other while continuing to interact strongly internally is an interesting theoretical problem. It is precisely this situation which arises in relativistic heavy ion collisions when the quark-gluon plasma that is formed as an intermediate state breaks up into many hadrons that fly apart and cease to interact among each other. The purpose of this article is to lay the groundwork for a (highly simplified and schematic) model that captures the essence of such a multifragmentation process at the quantum level\footnote{The highly complex initial state $|\Phi_{\rm i}\rangle$ of a relativistic heavy ion collision has nearly zero entropy. In the quark-gluon Fock space (parton) basis, this state is characterized by a density matrix with two blocks describing the two ground state nuclei approaching each other. This bipartite initial state evolves by a unitary transformation into a highly entangled final state $|\Phi_{\rm f}\rangle$ characterized by an even more complex density matrix in the parton basis that no longer has a block structure. Eventually, this final  state is projected onto hadron states by experimental measurements that identify the asymptotic eigenstates of the many-parton system (hadrons). This happens at a time of order $10^{-9}~{\rm s}$, long after the duration of the nuclear reaction proper which is of order $10^{-23}~{\rm s}$. The interaction of the hadrons with the detector leads to their decoherence, which is commonly interpreted as entropy production during the initial collision process, but is more precisely described as the measurement of the entanglement entropy present in the final hadron state.}. The results however hold generally for any holographic conformal field theory (CFT). In fact previous holographic calculations for single \cite{shimaji_holographic_2019} and double splits \cite{caputa_double_2019} qualitatively reproduce the results for non-holographic CFTs \cite{Calabrese_2007} and therefore we optimistically conjecture that our results hold for non-holographic CFTs as well.

We now turn to the model system that encapsulates such a break-up process. To wit, we study the sudden splitting of the ground state of a $(1+1)$-dimensional CFT on the infinite line into separate non-interacting segments. A more complete analogy to the situation described above would be obtained by starting from a highly excited pure quantum state on a finite interval with appropriate boundary conditions and its evolution under the fragmentation of the finite interval into smaller ones. We leave this more involved study for future work. As a first step, we here are concerned with the development of an approach that can capture the fragmentation of the infinite line into an arbitrary number of segments. The main result of this work is the application of this novel approach to the case of three segments by a double split, which reproduces earlier results obtained with a different conformal map (which we review) \cite{caputa_double_2019}.

\subsection{Splitting Quenches Review}

Calabrese and Cardy in a series of papers on entanglement entropy in quantum field theory \cite{Calabrese_2007,Calabrese_2004,Calabrese_2005,Calabrese_2009,calabrese_entanglement_2005,CARDY1989581,Cardy_2004} introduced the notion of a joining quench \cite{Calabrese_2007} , where two Boundary Conformal Field Theories [BCFTs] living on a half-line are joined at $t=0$ to form a boundaryless CFT living on the full line. Our work is concerned with the inverse process: A CFT living on the full line is split at $t=0$ into two or more BCFTs. This process is known as a splitting quench \cite{shimaji_holographic_2019}. 

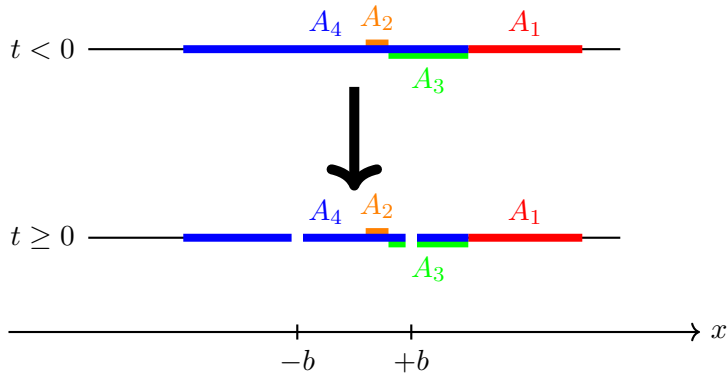
\begin{figure}
\centering
    
    



\tikzmath{\l=3.5;\h=2.5;\b=0.75;\c=0.075;\a=1.5;\aa=3;\k=0.15;\kk=0.45;\f=0.45;\ff=1.5;\m=-2.25;\mm=1.5;\off=0.075;}
\scalebox{1}{
\begin{tikzpicture}
    \draw[thick] ($(-\l,0)$)node[left=1]{$t<0$} -- ($(\l,0)$);
    \draw[thick] ($(-\l,-\h)$)node[left=1]{$t\geq0$} -- ($(\l,-\h)$);
    
    \draw[line width=3pt, red] ($(\a,0)$) -- ($(\aa,0)$)node[pos=0.5, above]{$A_1$};
    \draw[line width=3pt, orange] ($(\k,\off)$) -- ($(\kk,\off)$)node[pos=0.5, above]{$A_2$};
    \draw[line width=3pt, green] ($(\f,-\off)$) -- ($(\ff,-\off)$)node[pos=0.5, below]{$A_3$};
    \draw[line width=3pt, blue] ($(\m,0)$) -- ($(\mm,0)$)node[pos=0.5, above]{$A_4$};
    
    \draw[line width=3pt, red] ($(\a,-\h)$) -- ($(\aa,-\h)$)node[pos=0.5, above]{$A_1$};
    \draw[line width=3pt, orange] ($(\k,-\h+\off)$) -- ($(\kk,-\h+\off)$)node[pos=0.5, above]{$A_2$};
    \draw[line width=3pt, green] ($(\f,-\h-\off)$) -- ($(\ff,-\h-\off)$)node[pos=0.5, below]{$A_3$};
    \draw[line width=3pt, blue] ($(\m,-\h)$) -- ($(\mm,-\h)$)node[pos=0.5, above]{$A_4$};
    
    \draw[->, line width=0.8ex] ($(0,-0.2*\h)$) -- ($(0,-0.75*\h)$);

    \draw[line width=10pt, white] ($(\b-\c,-\h)$) -- ($(\b+\c,-\h)$);
    \draw[line width=10pt, white] ($(-\b-\c,-\h)$) -- ($(-\b+\c,-\h)$);

    \draw[thick, ->] ($(-1.3*\l,-1.5*\h)$) -- ($(1.3*\l,-1.5*\h)$)node[right]{$x$};
    \draw[thick] ($(-\b,-1.5*\h)+(0,-0.1)$)node[below]{$-b$} -- ($(-\b,-1.5*\h)+(0,0.1)$);
    \draw[thick] ($(\b,-1.5*\h)+(0,-0.1)$)node[below]{$+b$} -- ($(\b,-1.5*\h)+(0,0.1)$);
\end{tikzpicture}
}
\caption{Schematic visualization of the double splitting quench in 1+1D CFT. At time $t=0$, the system is cut at positions $x=\pm b$ yielding three distinct spatial regions for times $t\geq0$. Four qualitatively different types of subsystems can be identified: $A_1$ is located to the right of both cuts, $A_2$ lies in between the two cuts, $A_3$ is itself split by one cut and $A_4$ includes both cuts.}\label{fig:splitting_schematic}
\end{figure}
This term arises from the fact that just like a typical quantum quench the ground state of the Hamiltonian is prepared and at $t=0$ the Hamiltonian is instantaneously changed. Here, the only change is that the support of the Hamiltonian is not on the full line but two half-lines (or various subsets of the line for more than one split). In effect the line is split into multiple line segments at $t=0$. \autoref{fig:splitting_schematic} shows schematically the procedure of splitting for two cuts at spatial positions $x=\pm b$ at time $t=0$. For $t\geq0$, the Hamiltonian supports three distinct spatial regions $x\in[-\infty,-b)$, $x\in(-b,+b)$ and $x\in(+b,+\infty]$. We maintain conformal symmetry by holding that boundary conditions are chosen such that the conformal symmetry is broken from SO(2,2) to SO(2,1).\footnote{We will not comment further on this highly constrained choice as it is well known that it only enters into the holographic entanglement entropy as a constant term which shifts the transition time between the connected and disconnected geodesics \cite{caputa_double_2019,shimaji_holographic_2019}.}

\begin{figure}[htb]
\centering
\tikzmath{\taumax=3;\tmax=3;\l=1.5;\a=0.8;\ax=0.8;}
\tdplotsetmaincoords{70}{40}
\begin{tikzpicture}[tdplot_main_coords]
    \path[draw, fill=cyan!50!gray, semitransparent] ($(-\l,0,0)$) -- ($(-\l,-\taumax,0)$) -- ($(\l,-\taumax,0)$) -- ($(\l,0,0)$) --cycle;
    \node at ($(\l,-\taumax/2,0)$) [below right] {$\tau$};
    \node at ($(0,-\taumax,0)$) [below left] {$-\infty$};
    \path[draw, fill=cyan!50!gray, semitransparent] ($(-\l,0,0)$) -- ($(-\l,0,\tmax)$) -- ($(\l,0,\tmax)$) -- ($(\l,0,0)$) --cycle;
    \node at ($(\l,0,\tmax/2)$) [right] {$t$};
    \draw[thick] ($(0,-\a,0)$)node[below] {$-a$}--($(0,0,0)$);
    \draw[thick] ($(0,0,0)$)--($(0,0,\tmax)$);

    \coordinate[] (o) at ($(2*\l,0,0)$);
    \draw[thick, ->] (o) -- ($(o)+(\ax,0,0)$)node[right]{$x$};
    \draw[thick, ->] (o) -- ($(o)+(0,\ax,0)$)node[above]{$\tau$};
    \draw[thick, ->] (o) -- ($(o)+(0,0,\ax)$)node[above]{$t$};
\end{tikzpicture}
\caption{The world sheet for a splitting quench in imaginary and real time.}\label{DM_worldsheet}
\end{figure}
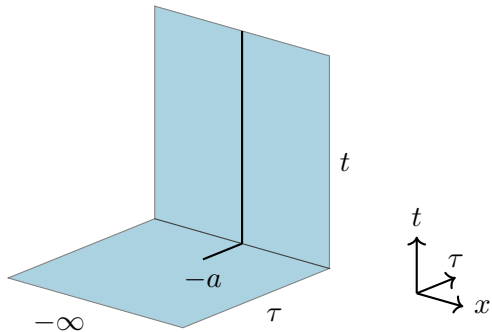

A natural question to ask is how the entanglement entropy of a given subsystem evolves as a function of time? Given that the segments were initially in a pure state and cease to interact after the quench, one should expect non-trivial dynamics of the entanglement entropy between different subsystems. This calculation can be accomplished in the worldsheet formalism by evolving in an infinite amount of Euclidean time to project onto the ground state and then evolving in Lorentzian time after the split. However, Calabrese and Cardy \cite{Calabrese_2005} realized that this process is singular and the joining/splitting operation must be extended an infinitesimal amount $a$ in Euclidean time,a s depicted in \autoref{DM_worldsheet}, to avoid divergences. The regulator should then be taken to 0 at the end of the calculation, a fact we make frequent use of in order to calculate in the small $a$ limit.

\begin{figure}[htb]
\centering
\tikzmath{\taumax=3;\tmax=3;\l=1;\a=1;}
\tdplotsetmaincoords{70}{40}
\scalebox{0.8}{
\begin{tikzpicture}[tdplot_main_coords]
    \path[draw, fill=cyan!50!gray, semitransparent] ($(-\l,\taumax,0)$) -- ($(-\l,\taumax,\tmax)$) -- ($(\l,\taumax,\tmax)$) -- ($(\l,\taumax,0)$) --cycle;
    \path[draw, fill=cyan!50!gray, semitransparent] ($(-\l,\taumax,\tmax)$) -- ($(-\l,0,\tmax)$) -- ($(\l,0,\tmax)$) -- ($(\l,\taumax,\tmax)$) --cycle;
    \path[draw, fill=cyan!50!gray, semitransparent] ($(-\l,0,0)$) -- ($(-\l,-\taumax,0)$) -- ($(\l,-\taumax,0)$) -- ($(\l,0,0)$) --cycle;
    \node at ($(\l,-\taumax,0)$) [below right] {$-a-it$};
    \node at ($(\l,0,0)$) [below right] {$-it$};
    \node at ($(0,-\taumax,0)$) [below left] {$|\psi_0\rangle$};
    \path[draw, fill=cyan!50!gray, semitransparent] ($(-\l,0,0)$) -- ($(-\l,0,\tmax)$) -- ($(\l,0,\tmax)$) -- ($(\l,0,0)$) --cycle;
    \node at ($(0,0,\tmax)$) [below] {$|\psi^{''}\rangle$};
    \node at ($(0,0,\tmax)$) [above right] {$|\psi^{'}\rangle$};
    \node at ($(-\l,0,\tmax)$) [left] {$0$};
    \node at ($(\l,\taumax,\tmax)$) [right] {$a$};
    \node at ($(\l,\taumax,0)$) [right] {$a-it$};
    \node at ($(0,\taumax,0)$) [below] {$\langle\psi_0|$};
\end{tikzpicture}
}
\tikzmath{\taumax=5;\tmax=3;\l=1.2;\a=3;\ax=0.8;}
\tdplotsetmaincoords{70}{40}
\scalebox{0.8}{
\begin{tikzpicture}[tdplot_main_coords]
    \path[draw, fill=cyan!50!gray, semitransparent] ($(-\l,\a,0)$) -- ($(-\l,\taumax,0)$) -- ($(\l,\taumax,0)$) -- ($(\l,\a,0)$) --cycle;
    \path[draw, fill=cyan!50!gray, semitransparent] ($(-\l,\a,0)$) -- ($(-\l,\a,\tmax)$) -- ($(\l,\a,\tmax)$) -- ($(\l,\a,0)$) --cycle;
    \path[draw, fill=cyan!50!gray, semitransparent] ($(-\l,\a,\tmax)$) -- ($(-\l,0,\tmax)$) -- ($(\l,0,\tmax)$) -- ($(\l,\a,\tmax)$) --cycle;
    \path[draw, fill=cyan!50!gray, semitransparent] ($(-\l,0,0)$) -- ($(-\l,-\a,0)$) -- ($(\l,-\a,0)$) -- ($(\l,0,0)$) --cycle;
    \node at ($(\l,-\a,0)$) [below right] {$-a-it$};
    \node at ($(\l,0,0)$) [below right] {$-it$};
    \node at ($(0,-\a,0)$) [above] {$|0\rangle$};
    \path[draw, fill=cyan!50!gray, semitransparent] ($(-\l,0,0)$) -- ($(-\l,0,\tmax)$) -- ($(\l,0,\tmax)$) -- ($(\l,0,0)$) --cycle;
    \path[draw, fill=cyan!50!gray, semitransparent] ($(-\l,-\a,0)$) -- ($(-\l,-\taumax,0)$) -- ($(\l,-\taumax,0)$) -- ($(\l,-\a,0)$) --cycle;
    \draw[thick] ($(0,-\a,0)$)--($(0,0,0)$);
    \draw[thick] ($(0,\a,\tmax)$)--($(0,0,\tmax)$);
    \draw[thick] ($(0,0,0)$)--($(0,0,\tmax)$);
    \draw[thick] ($(0,\a,0)$)--($(0,\a,\tmax)$);
    \node at ($(0,0,\tmax)$) [below left] {$|\psi^{''}\rangle$};
    \node at ($(0,0,\tmax)$) [above] {$|\psi^{'}\rangle$};
    \node at ($(-\l,0,\tmax)$) [left] {$0$};
    \node at ($(\l,\a,\tmax)$) [right] {$a$};
    \node at ($(\l,\a,0)$) [below right] {$a-it$};
    \node at ($(\l,\taumax,0)$) [below right] {$\infty-it$};
    \node at ($(\l,-\taumax,0)$) [below right] {$-\infty-it$};
    \node at ($(0,\a,0)$) [below] {$\langle0|$};

    \coordinate[] (o) at ($(-4*\l,0,0)$);
    \draw[thick, ->] (o) -- ($(o)+(\ax,0,0)$)node[right]{$x$};
    \draw[thick, ->] (o) -- ($(o)+(0,\ax,0)$)node[above]{$\tau$};
    \draw[thick, ->] (o) -- ($(o)+(0,0,\ax)$)node[left]{$t$};
\end{tikzpicture}
}
\caption{Left: The world sheet for \ref{GlobQuench} where $|\psi_0\rangle$ is an arbitrary initial state. 
Right: The world sheet for the splitting quench.}
\label{fig:Worldsheets}
\end{figure}
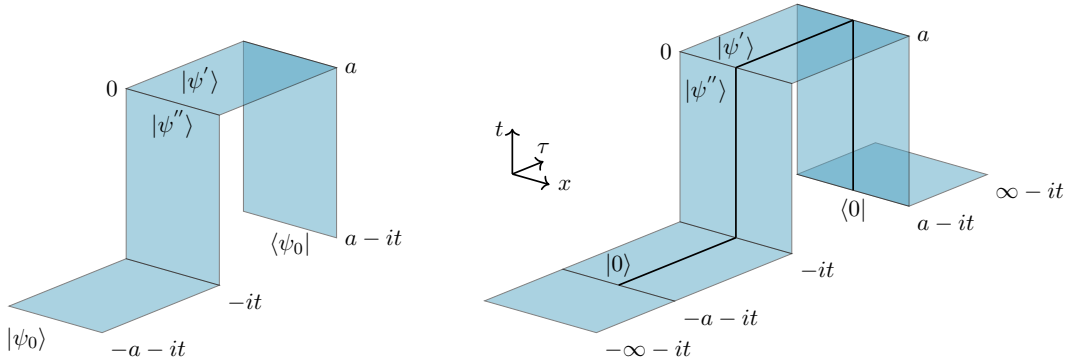

To get the entanglement entropy as a function of time we need the density matrix $\rho(t)=\ket{0(t)}\bra{0(t)}$ which can then be analytically continued to find its evolution in Lorentzian time.
The density matrix with this regulator is then as follows (where the $a\rightarrow 0$ limit recovers the standard density matrix)
\be 
\label{GlobQuench}
\braket{\psi''(x'')|\rho(t)|\psi'(x')}
=Z^{-1}\braket{\psi''(x'')|e^{-itH-aH}|\psi_0}\braket{\psi_0|e^{+itH-aH}|\psi'(x')} ,
\ee
which can be represented by the worldsheet shown in the left panel of \autoref{fig:Worldsheets}.

Now consider a splitting quench where $\ket{\psi_0}$ is the ground state of the CFT on a line (achieved by evolving an infinite amount in Euclidean time) and we time evolve with two half-lines (as the support of the Hamiltonian changes). Eq. (\ref{GlobQuench}) is then visualized by the right part of \autoref{fig:Worldsheets}. If we consider $\braket{\psi''(x'') |\rho(0)|\psi'(x')}$ we are left with the euclidean plane with a slit in it. Introducing a complex coordinate $z=x+it$ we see that the cut extends over $[-ia,ia]$ and that the junction between $\bra{\psi''}$ and $\ket{\psi'}$ is on the $x$ axis.

When we want to compute the reduced density matrix we sew the worldsheet together along the complement of the region under consideration. e.g. (if we wish to compute $\rho_A$ with $A=[-l,l]$ then we would sew from $[-\infty,-l]$ and $[l,\infty]$ leaving a cut along the region A). It was shown in \cite{Calabrese_2004} that using the replica trick to compute this reduced density matrix is equivalent to calculating the correlation function of primary operators placed at the endpoints of $A$. Later, \cite{Ryu_2006a} showed that one can compute this 2-point function of twist operators (the appropriate primary operators) holographically by considering the length of the geodesic in the bulk between the two endpoints of subsystem $A$. Therefore if we want to determine the entanglement entropy of any subsystem it suffices to consider the path-integral on the plane with a cut from $[-ia,ia]$, viz. 
\begin{align}
\braket{\psi''(x'')|\rho_A|\psi'(x')}
&=Z^{-1}{\rm Tr}_B\braket{\psi''(x'')|e^{-aH}|0}\braket{0|e^{-aH}|\psi'(x')}
\\
&=\braket{0|e^{-aH} \Phi(x_1+i\tau) \Phi(x_2+i\tau) e^{-aH}|0},
\end{align}
and then to follow the Holographic Entanglement Entropy (HEE) prescription \cite{Ryu_2006a} on the Euclidean world sheet by calculating the shortest geodesic homologous to the line segment A.

What if we wish to split the line at two points? That is to split our CFT on the line into two rays and one line segment. A calculation of the entanglement entropy over time has been carried out in \cite{caputa_double_2019}, in this technical note we reproduce the result with two additional maps which allow more easily for generalization to the case of more than two cuts.

It is quite natural to assume that the calculation gets significantly harder as more cuts are added. When we have two cuts we have two parameters: the size of the cuts and their distance. It turns out that the space of conformally inequivalent worldsheets only depends on the ratio of these two parameters, however as we add more cuts the situation gets more complex. The world sheet under consideration, the complex plane with $n$ cuts, is the Riemann surface of genus 0 with $n$ cuts. It is a well-known fact\footnote{To see this consider that the Schottky double (that is the branched double cover) of the Riemann surface with $n$ boundaries and genus 0 is the genus $g=n-1$ surface. The size of the moduli space of the genus $g$ surface is of complex dimension $3g-3$ and real dimension $6g-6$.} that the space of conformally equivalent such surfaces is of real dimension $6n-12$. However, as we will show, since all cuts are of the same size and the regulator is taken to 0, we are in a specific corner of moduli space where many things simplify. The same issue does not arise when considering multiple joining quenches as the splits ``intersect'' at the point at infinity, meaning the space is simply connected and the Riemann mapping theorem is sufficient.

\subsection{Structure of Our Manuscript}
We present three ways of calculating the entanglement entropy of a world sheet with two cuts, two of which are novel. First, for the convenience of the reader, we reproduce the modified theta function map of \cite{caputa_double_2019}. Then we introduce the Abel-Jacobi map, which is the inverse of the modified theta function map of \cite{caputa_double_2019}, showing that it simplifies calculations dramatically. Finally, we introduce our full prescription which involves constructing the holographic dual of the Schottky uniformization of the world sheet with two cuts. We do this by using an explicit conformal map based on the Schottky-Klein prime function \cite{Crowdy2013} on the Schottky uniformization domain ($n$ circles cut out of the unit disk, where $n$ is the number of cuts). All three holographic entanglement entropy calculations give the same result which provides a check of our prescription.\\
In Appendix A we show that the various forms of the BTZ which appear in our manuscript are equivalent and are just due to different slicing in the bulk coordinate. In Appendix B we give a review of the explicit calculation of the geodesic in the BTZ black hole.

\subsection{Summary of Results}

\begin{figure}
    \centering
    \includegraphics[width=\linewidth]{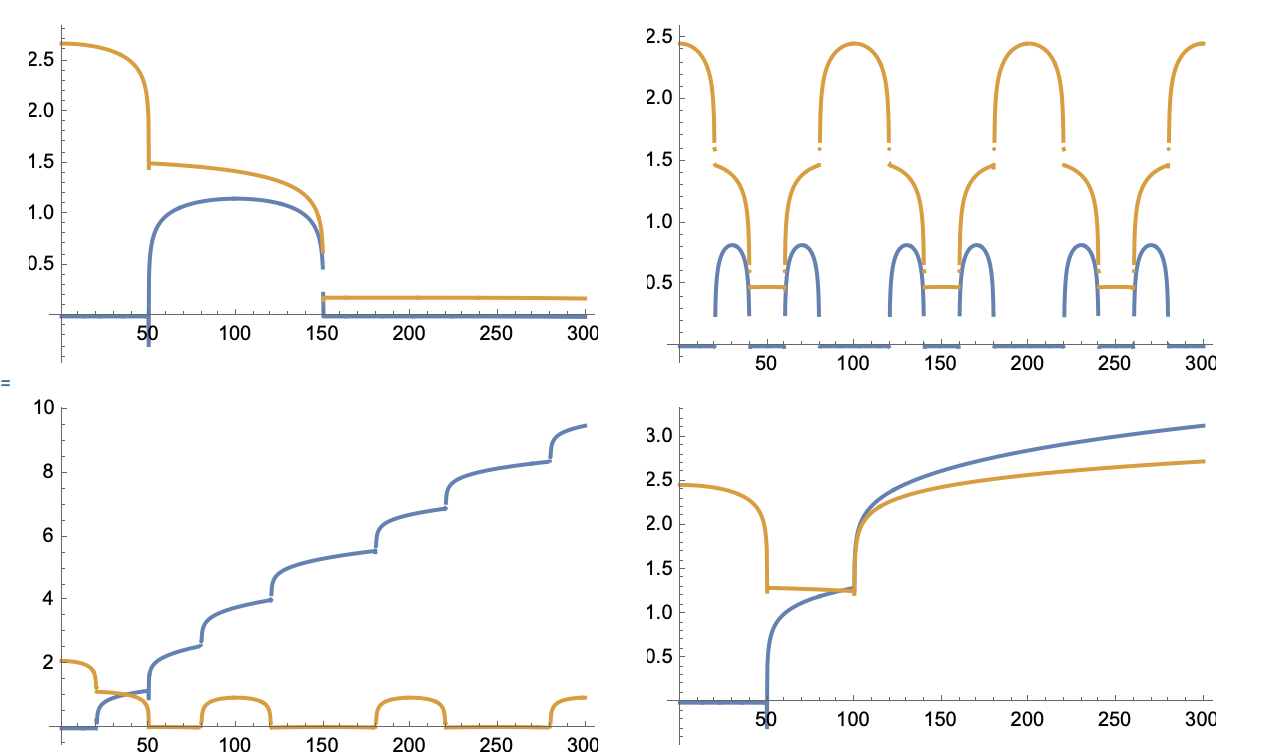}
    \caption{Numerical calculation of the holographic entanglement entropy growth $\Delta S_A$ for the subsystems $A_1$, $A_2$, $A_3$ and $A_4$ with parameters $b=50$, $a=0.05$ and $S_{{\rm bdy},1}=S_{{\rm bdy},2}=0$ using Schottky uniformization. The blue and orange lines are the connected and disconnected entropy growth respectively.}
    \label{fig:S_SU}
\end{figure}

Comparison of the numerical calculations of holographic entanglement entropy for four qualitatively different subsystems using the three different conformal map approaches (\autoref{fig:S_inv}, \autoref{fig:S_AJ} and \autoref{fig:S_SU}) yields perfect agreement. This confirms the equivalence of the modified theta function map with the two other approaches presented in this work. We interpret the worldsheet with cuts as the projection of an appropriate (hyper)elliptic curve onto the complex plane. This allows us to make use of the theory of Riemann surfaces (see \cite{baker1995abelian,farkas2012riemann,mumford1983tata,fay1973theta,Dubrovin,Grava} for well-known reviews). We typically take the Schottky double (a branched double cover over the cuts) in order to deal with the simpler case of compact Riemann surfaces – the original surface with boundary can be recovered via involution.

The Abel-Jacobi map is a map from a compact Riemann surface to its Jacobian variety (a variety constructed from the periods of the Riemann surface on its A and B cycles). For the case of an elliptic curve (the Schottky double of the two slit plane) the Jacobian variety is also a torus and therefore the map is a conformal transformation from the branched covering of $\mathbb{C}$ to the doubly identified rectangle (see \autoref{fig:3reps}). We show that this provides the inverse of the modified theta function presented in \cite{caputa_double_2019}. It has the benefit of easily generalizing to multiple cuts and being simpler to calculate.

\begin{figure}
    \centering
    \includegraphics[width=0.5\linewidth]{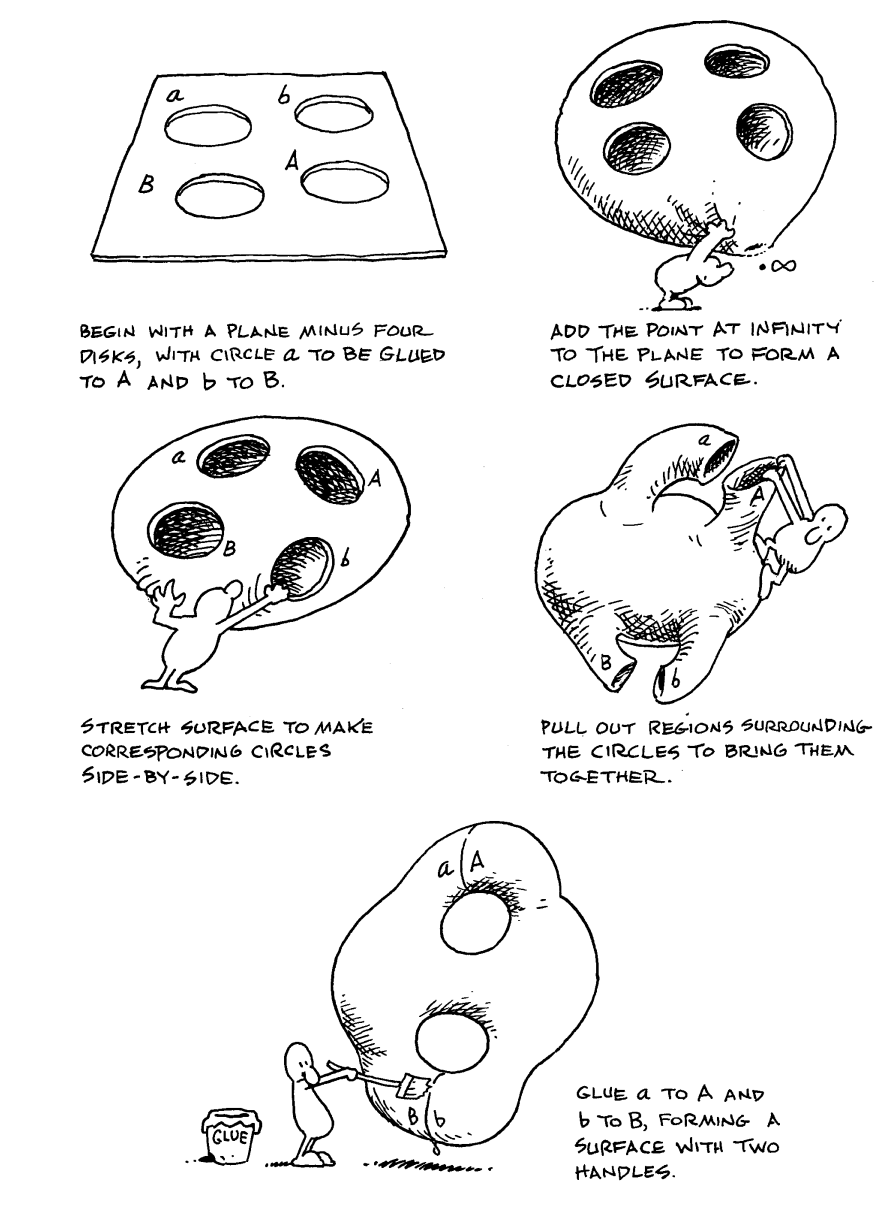}
    \caption{An example of the Schottky uniformization of the genus 2 surface from Indra's Pearls by Mumford, et al. \cite{Mumford_Series_Wright_2002}.}
    \label{fig:Uniformization}
\end{figure}

Riemann surfaces of genus larger than 1 (equivalently of more than 2 boundaries) have moduli, parameters of the surface that give coordinates on the space of conformally inequivalent Riemann surfaces of the same genus (number of boundaries). It is convenient to have a description of our Riemann surfaces in terms of a fundamental domain that depends on the moduli. This is known as uniformization. The best known way to uniformize compact Riemann surfaces is to take the quotient of the upper-half plane with a Fuchsian group (a subgroup of $SL(2, \mathbb{R})$). This is known as Fuchsian uniformization and will be familiar to those who have calculated string amplitudes. Another method is due to Schottky, where one uniformizes a Riemann surface with a fundamental domain that is $g$ circles cut out of the unit disk. Then the Schottky double is obtained by mirroring these disks across the unit disk, and the compact Riemann surface is constructed by identifying the circles with their mirror images (viz. taking the quotient with a subgroup of $SL(2, \mathbb{C})$. See \autoref{fig:Uniformization} for a artist's depiction). The holographic dual of the Schottky uniformization of a compact Riemann surface was worked out in \cite{Krasnov2000}. We make use of new results on conformal mappings \cite{crowdy2007computing,Crowdy_2008,crowdy2012conformal,Crowdy2013,crowdy2020solving} to find the inverse of the map from the plane with multiple cuts to its Schottky uniformization. We then solve for the inverse map numerically. We are able to obtain an explicit expression for this map in the $a\rightarrow0$ limit which allows us to find the metric and therefore the geodesic lengths.

This novel prescription – mapping the worldsheet to its Schottky uniformization and then constructing the holographic dual – is shown here to match earlier methods \cite{caputa_double_2019}. The explicit calculation of entanglement entropy as a function of time for more than two splits will be explored in a forthcoming paper.

\section{The 3 Conformal Maps}

\begin{figure}
    \centering
    \tikzmath{\l=3.85;\a=0.6;\b=0.8;\d=2;\o=0.1;\n=0.9;}
\tdplotsetmaincoords{70}{20}
\scalebox{1}{
\begin{tikzpicture}[tdplot_main_coords]
    \coordinate (pmbma) at ($(-\b,-\a, 0)$);
    \coordinate (pmbpa) at ($(-\b,\a, 0)$);
    \coordinate (ppbma) at ($(\b,-\a, 0)$);
    \coordinate (ppbpa) at ($(\b,\a, 0)$);
    
    \coordinate (mmbma) at ($(-\b,-\a, -\d)$);
    \coordinate (mmbpa) at ($(-\b,\a, -\d)$);
    \coordinate (mpbma) at ($(\b,-\a, -\d)$);
    \coordinate (mpbpa) at ($(\b,\a, -\d)$);
    
    \path[fill=blue!50!gray, semitransparent] ($(-\l/2, -\l/2, -\d)$) -- ($(\l/2, -\l/2, -\d)$) -- ($(\l/2, \l/2, -\d)$) -- ($(-\l/2, \l/2, -\d)$) --cycle;
    \draw[red, ultra thick] (mmbma) -- (mmbpa);
    \draw[green, ultra thick] (mpbma) -- (mpbpa);

    \path[fill=gray!80, semitransparent] (mmbma)--(pmbma)--(pmbpa)--(mmbpa) --cycle;
    \path[fill=gray!80, semitransparent] (mpbma)--(ppbma)--(ppbpa)--(mpbpa) --cycle;

    \path[fill=cyan!50!gray, semitransparent] ($(-\l/2, -\l/2, 0)$) -- ($(\l/2, -\l/2, 0)$) -- ($(\l/2, \l/2, 0)$) -- ($(-\l/2, \l/2, 0)$) --cycle;
    \draw[red, ultra thick] (pmbma) -- (pmbpa);
    \draw[green, ultra thick] (ppbma) -- (ppbpa);

    \coordinate (z) at ($(1.5*\b,2*\b,0)$);
    
    \draw[yellow, thick] ($(-\l/2, 0, 0)$) -- ($(-\b, 0, 0)$) --($(-\b, 0, -\d)$) --($(-\l/2, 0, -\d)$);
    \path[draw, orange, thick] ($(\b,\o,0)$)--($(-\b,\o,0)$)--($(-\b,\o,-\d)$)--($(\b,\o,-\d)$)--cycle;
    \draw[blue, thick] ($(\l/2, 0, 0)$) -- ($(\b, 0, 0)$) --($(\b, 0, -\d)$) --($(\l/2, 0, -\d)$);

    \draw[] ($(\l/2, \l/2, 0)+(-\n,0,0)$)--($(\l/2, \l/2, 0)+(-\n,-\n,0)$)--($(\l/2, \l/2, 0)+(0,-\n,0)$) node[pos=0.7, above ] {$w$};
    
\end{tikzpicture}
}
    \tikzmath{\l=2.5;\h=2.1;\b=1;\a=1;\n=0.5;\off=0.2;}
\scalebox{1}{
\begin{tikzpicture}
    \path[fill=cyan!50, opacity=0.5] ($(-\l/2,-\h/2)$)--($(\l/2,-\h/2)$)--($(\l/2,\h/2)$)--($(-\l/2,\h/2)$)--cycle;
    \path[fill=blue!50, opacity=0.5] ($(-\l/2,-\h)$)--($(\l/2,-\h)$)--($(\l/2,-\h/2)$)--($(-\l/2,-\h/2)$)--cycle;
    \path[fill=blue!50, opacity=0.5] ($(-\l/2,\h/2)$)--($(\l/2,\h/2)$)--($(\l/2,\h)$)--($(-\l/2,\h)$)--cycle;
    \draw[black, ultra thick, dashed] ($(-\l/2,-\h)$)--($(\l/2,-\h)$);
    \draw[black, ultra thick, dashed] ($(-\l/2,\h)$)--($(\l/2,\h)$);
    
    \draw[red, ultra thick] ($(-\l/2,\h/2)$)node[left]{$\color{black}\frac{i\tau}{4}$}--($(\l/2,\h/2)$);
    \draw[green, ultra thick] ($(-\l/2,-\h/2)$)node[left]{$\color{black}-\frac{i\tau}{4}$}--($(\l/2,-\h/2)$);
    \draw[yellow, ultra thick,->] ($(-\l/2,0)$)node[left]{$\color{black}0$}--($(-\l/2,\h)$)node[left]{$\color{black}\frac{i\tau}{2}$}node [pos=0.75, right] {};
    \draw[yellow, ultra thick,->, dashed] ($(\l/2,0)$)--($(\l/2,\h)$);
    \draw[blue, ultra thick,->] ($(-\l/2,0)$)--($(-\l/2,-\h)$)node[left]{$\color{black}-\frac{i\tau}{2}$}node [pos=0.75, right] {};
    \draw[blue, ultra thick,->, dashed] ($(\l/2,0)$)--($(\l/2,-\h)$);
    \node[right] at  ($(\l/2,0)$) {$\color{black}1$};
    \draw[orange, ultra thick,->] ($(0,0)$)node[left]{$\color{black}\frac{1}{2}$}--($(0,\h)$);
    \draw[orange, ultra thick,->] ($(0,-\h)$)--($(0,0)$)node [pos=0.8, right] {};
    
    \draw[] ($(\l/2, \h)+(-\n,0)-(\off,\off)$)--($(\l/2, \h)+(-\n,-\n)-(\off,\off)$)--($(\l/2, \h)+(0,-\n)-(\off,\off)$) node[midway, above] {$\eta$};

    \node [cross, label=above:{}] (a) at (\l/2,0) {};
    \node [cross, label=above:{}] (b) at (-\l/2,0) {};
    \node [cross, label=above:{}] (c) at (\l/2,\h) {};
    \node [cross, label=above:{}] (d) at (-\l/2,\h) {};
    \node [cross, label=above:{}] (c) at (\l/2,-\h) {};
    \node [cross, label=above:{}] (d) at (-\l/2,-\h) {};
\end{tikzpicture}
}
    \tikzmath{\r=0.7;}
\scalebox{1}{
\begin{tikzpicture}
    \fill [cyan!50!gray, semitransparent,even odd rule] (0,0) circle[radius=\r] circle[radius=2*\r];
    \fill [blue!50!gray, fill opacity=0.5,even odd rule] (0,0) circle[radius=2*\r] circle[radius=3*\r];
    \draw [name path=B, ultra thick, draw=green](0,0) circle (2*\r);
    \draw [name path=A, ultra thick, draw=red, dashed](0,0) circle (\r);
    \draw [name path=C, ultra thick, draw=red, dashed](0,0) circle (3*\r);
    
    \draw [orange, ultra thick] (-\r,0) node [right] {\color{black}$\rho$} --(-3*\r,0)node [left] {\color{black}$\frac{1}{\rho}$} node [pos=0.5,above right] {\color{black}$1$};
    \draw [yellow, ultra thick] (\r,0)--(1.5*\r,0);
    \draw [blue, ultra thick] (1.5*\r,0)--(2.5*\r,0);
    \draw [yellow, ultra thick] (2.5*\r,0)--(3*\r,0);
    \node [cross, label=above:{$\alpha$}] (a) at (1.5*\r,0) {};
    \node [cross, label=above:{$\frac{1}{\alpha}$}] (a) at (2.5*\r,0) {};

    \draw [thick](0,2.8*\r) --(0,2.2*\r)--(0.6*\r,2.2*\r)node [pos=0.8, above] {\color{black}$re^{i\theta}$};
\end{tikzpicture}
}
    \caption{3 representations of the torus from left to right: branched covering of $\mathbb{C}$, doubly periodic rectangle, identified annulus. Dotted lines signify identification. X's mark points excluded from the space (the preimage of the points at infinity on the upper and lower sheet).}
    \label{fig:3reps}
\end{figure}
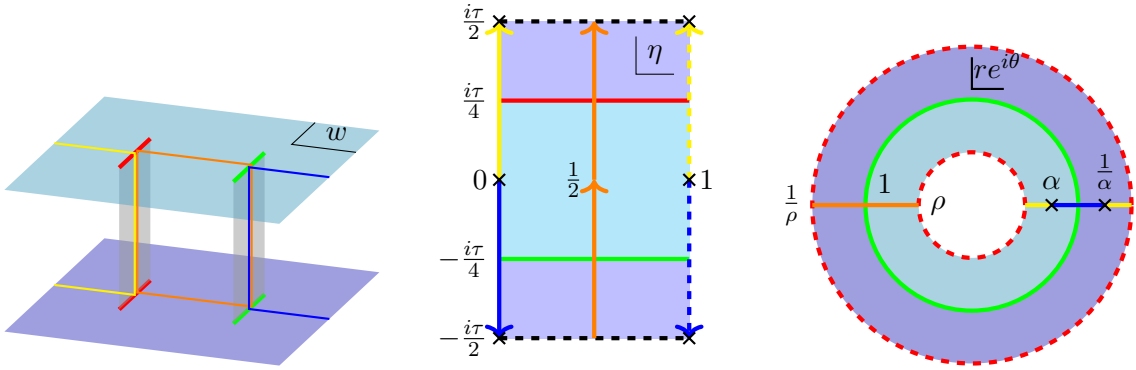
The starting point of our technical approach is a world sheet with two cuts in it. Since our field theory is conformally invariant, the physics will be invariant under conformal transformations. The name of the game is to find a conformal mapping to a domain that greatly simplifies calculation. In the case of one cut, Calabrese and Cardy used the mapping to the upper half plane \cite{calabrese_entanglement_2005}, and for two cuts Caputa, et al. \cite{caputa_double_2019} numerically found the inverse of a truncation of a modified theta function in order to calculate the geodesics in the BTZ metric.

We will demonstrate two additional options, which allow for straightforward generalization to more than two cuts (to be explored in detail in a forthcoming paper). We interpret the world sheet with two slits as the projection of an elliptic curve onto the Riemann sphere and take the Schottky double to recover the typical ramified double cover presentation as depicted in the left of \autoref{fig:3reps}. By understanding the worldsheet in this way we can make use of the classical theory of (hyper)elliptic curves and more recent advances in conformal mappings of such domains \cite{Crowdy2013,crowdy2007computing,Crowdy_2008,crowdy2012conformal,crowdy2020solving}. Rather than numerically finding the inverse of a truncation of the theta function we use the Abel-Jacobi map which is the natural conformal map from an elliptic curve to its Jacobian variety (which in the case of a torus/elliptic curve is also a torus). This is the exact inverse of the map used in \cite{caputa_double_2019}. In addition we map the two slit plane to the annulus (its Schottky uniformization) by numerically solving for the inverse of a truncated version of the map from the annulus. This map is based on a special function, the Schottky-Klein Prime function, and the asymptotics become exact in the limit $a\rightarrow0$.

We chose to use $w=x+it,\bar{w}$ coordinates for the complex plane and $z$ as its AdS bulk coordinate, $\eta=\Xi+i\Upsilon,\bar{\eta}$ for the doubly periodic rectangle with bulk coordinate $\zeta$, and cylindrical coordinates for the identified annulus $r,\theta,\xi$

\subsection{Theta Function}
Caputa et al. \cite{caputa_double_2019} use a modified theta function to conformally map the rectangle with sides identified to the plane with two cuts. In this section we retrace their method.
The inverse conformal map is given by 
\begin{align}
    w(\eta)=b\left[K(\eta)+K\left(\eta+\frac{i\tau}{2}\right)+1\right]\label{eq:inverse_map_equation}
\end{align}
where the function $K$ is defined as
\begin{align}
    K(\eta)&=\frac{1}{i\pi}\partial_{\eta'}\theta_1(\eta',i\tau)|_{\eta=\eta'}\label{eq:K}\\
    \theta_1(\eta',i\tau)&=2e^{-\pi\tau/4}\sin{\pi\eta}\prod_{k=1}^\infty(1-e^{-2\pi k\tau})(1-e^{2\pi i\eta}e^{-2\pi k\tau})(1-e^{-2\pi i\eta}e^{-2\pi k\tau}).\label{eq:theta}
\end{align}
It maps the complex plane with two cuts at $\pm b$ and regulator $a$ described by the complex coordinate $w=x+i t$ bijectively to the rectangle $\left[0,1\right]\times(-i\tau/4,i\tau/4)$ with coordinate $\eta$ where the lines $\textrm{Im}(\eta)=\pm \tau/4$ correspond to the two cuts respectively. The lines $\textrm{Re}(\eta)=0,1$ are identified, thus yielding an annulus geometry of the mapped rectangle. The modular parameter $\tau$ depends on the ratio $b/a$ and in the case $b\gg a$ takes the form \cite{caputa_double_2019}
\begin{align}
    \tau\approx\frac{2}{\pi}\ln\frac{4b}{a}.
\end{align}
Thus, in the limit $a\rightarrow 0$, which we are interested in, we may assume $\tau\rightarrow \infty$ leading to major simplifications.\\
So far this has all been in the Euclidean framework. After Wick rotation the Lorentzian coordinates $w_\pm=x\pm t$ and their corresponding annulus coordinates $\eta_\pm$ are connected via
\begin{align}
        \begin{split}
            x-t&=b\left[K(\eta_-)+K\left(\eta_-+\frac{i\tau}{2}\right)+1\right]\\
            x+t&=b\left[\Bar{K}(\eta_+)+\Bar{K}\left(\eta_+-\frac{i\tau}{2}\right)+1\right].\label{eq:inverse_nup_num}
        \end{split}
    \end{align}
    
\subsection{Abel-Jacobi Map}\label{sec:AJ_map}

The inverse of the map used by Takayanagi et al. viz. the map from the plane with two slits to the rectangle with sides identified is the classical Abel-Jacobi map. First we take the Schottky double of the plane with two slits, that is the double cover where the slits are branch cuts that connect the two sheets. Then we can fix the canonical homology basis on said elliptic curve, that is we choose two cycles, an A cycle and a B cycle, in a standard manner such that any loop on our space can be described by a linear combination of the two cycles. Taking the integral of the fundamental differential along the A and B cycles provides a conformal map to the Jacobian variety associated to the curve. For the torus this is a holomorphic bijection and therefore we have a conformal mapping. In general a Riemann surface of genus $g$ maps to a $2g$ torus.

We interpret our two cuts on the plane as the projection onto $\mathbb{C}$ of the elliptic curve $y^2=(x-(b+ia))(x-(b-ia))(x-(-b+ia))(x-(-b-ia))$ where the branch cuts are $[-b-ia,-b+ia]$ and $[b-ia,b+ia]$.

\begin{figure}[H]
    \centering
    \includegraphics[width=0.8\linewidth]{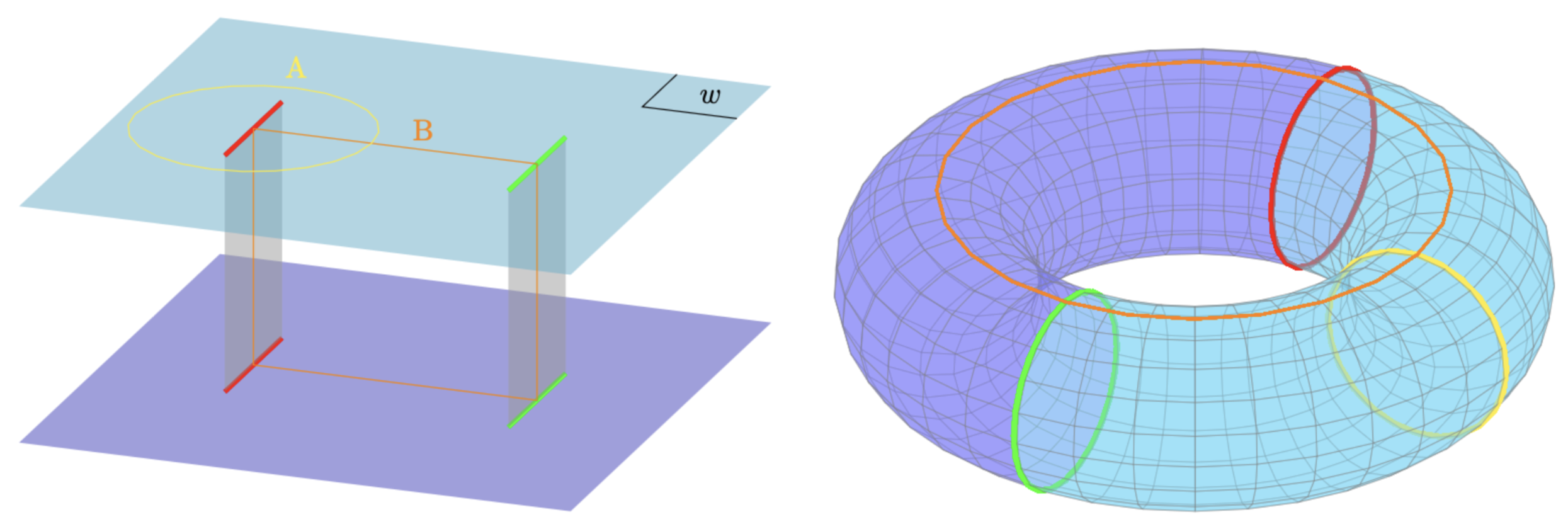}
    \caption{Schematic visualization of the Schottky doubled world sheet (left) and the conformally equivalent torus (right) embedded in three dimensions. The red and green lines represent the cuts at $-b$ and $+b$ respectively. Further, the A- and B cycle are shown as yellow and orange curve respectively.}
    \label{fig:Schottky_torus}
\end{figure}

The homology of our space is non-trivial (as it's topologically a torus) so we fix a canonical homology basis, choosing a loop around the left cut to be the A cycle and the loop between the sheets as the B cycle as is standard. This choice of cycles is visualized on the Schottky double and on the torus in \autoref{fig:Schottky_torus}.

There's a canonical differential associated to the elliptic curve
\be
\omega=\frac{dx}{y}=\frac{dx}{\sqrt{(x^2-(b+ia)^2)(x^2-(b-ia)^2)}}\label{eq:canonical_differential}
\ee
and we can integrate it over the A and B cycles to get two periods. These periods define a lattice in the complex plane that gives us a torus when opposite edges are identified\footnote{Since the Jacobian variety of the elliptic curve and the elliptic curve are both tori they are biholomorphic. We take pains to distinguish them as in the hyperelliptic case the Jacobian variety is a $2g$ torus rather than a genus-$g$ surface.}.
The Abel-Jacobi map is a conformal map from our original elliptic curve (the plane with two slits) to the Jacobian variety (the torus defined by the lattice). It is defined as a contour integral from a chosen base point to the point to be mapped of the canonical differential\footnote{This calculation is well-defined and easily computed by Mathematica's NIntegrate, however two simplifications are common and make this analytically doable. Typically one scales by the period of the A cycle so the lattice is defined by 1 and a complex (in our case imaginary) number $\tau=\frac{\int_B\omega}{\int_A\omega}$. In addition one may take advantage of the 3 DoF of the conformal symmetry to map the endpoints of the cuts to $0,1,\lambda,\infty$. In this case one gets that $\lambda=1+(\frac{b}{a})^2$ and $\tau=\frac{\int_B\omega}{\int_A\omega}=i\sqrt{\lambda}\frac{K(1-\lambda)}{K(1/\lambda)}$ where K is the elliptic K function. For the choice $a=.05,b=50$ this gives us $\tau\sim 5.280$}:
\begin{align}
\eta(w)=\int_{w_0}^w \omega=\int_{w_0}^w \frac{dw'}{\sqrt{(w'^2-(b+ia)^2)(w'^2-(b-ia)^2)}}
\end{align}
The choice of the base point results in a shift of the torus in the complex plane and therefore we choose $\infty$ for convenience.

\subsection{Schottky uniformization}

Since our curves depend on $6n-12$ parameters known as moduli, it is convenient to have a description of our Riemann surfaces in terms of a fundamental domain that depends on the moduli. We follow the method due to Schottky in which the fundamental domain consists of $g$ circles cut out of the unit disk. Then the Schottky double consists of the mirror image of these disks across the unit disk and the compact Riemann surface is constructed by identifying the circles with their mirror images (viz. modding out by a subgroup of $SL(2,\mathbb{C})$). Such a procedure naturally extends holographically \cite{Krasnov2000} as $SL(2,\mathbb{C})$ naturally extends to an isometry of $\mathbb{H}_3$ (Euclidean $AdS_3$). Hence, the appropriate holographic dual geometry is given by a quotient of the bulk by the Schottky group.

Here we use the conformal map from the fundamental Schottky domain to the plane with 2 slits to compute the holographic entanglement entropy. The map is constructed from the Schottky-Klein prime function in a method due to Crowdy and Marshall (a pedagogical introduction to which is found here \cite{crowdy2020solving}). The Schottky-Klein prime function has a presentation due to Baker \cite{baker1995abelian}
\begin{align}
\omega(z;\alpha)=(z-\alpha)\prod_{\vartheta\in\Theta}\frac{( z-\vartheta(\alpha))(\alpha-\vartheta(z))}{(\alpha-\vartheta(\alpha))(z-\vartheta(z))}
\end{align}
where $\alpha$ is the pre-image of the point at infinity and where $\vartheta$ are the elements of the Schottky Group $\Theta$, viz. compositions of the $n-1$ M\"{o}bius maps that send the circles in the unit disk to their reflections, excluding the identity map and inverse maps.\\
The appropriate conformal map is then:
\begin{align}
w(r,\theta)=\frac{1}{10\sqrt{(1+\alpha^{-2})(1+\bar{\alpha}^{-2})}\rho}\lb\frac{1}{2\bar{\alpha}}-\frac{1}{\bar{\alpha}^2}\partial_{\bar{\alpha}^{-1}} \ln(\omega(re^{i\theta};\bar{\alpha}^{-1}))+\frac{3}{\alpha}-\partial_\alpha \ln(\omega(re^{i\theta};\alpha))\rb
\end{align}
Where $\alpha$ is chosen such that the two slits are of the same size giving the original plane with two cuts. In the $a\rightarrow0$ limit this expression drastically simplifies\footnote{We can take the $a\rightarrow 0$ limit as the inner radius of the annulus $\rho$ is a function of $a$ and $b$ and the $a\rightarrow0$ limit corresponds to the $\rho\rightarrow0$ limit. See our forthcoming paper for the asymptotics for the multiple cut case where we show the infinite product truncates to the generators in the $\rho\rightarrow0$ limit.}.
\begin{align}
\lim_{a\rightarrow0}w(r,\theta)=\frac{1}{10\sqrt{(1+\alpha^{-2})(1+\bar{\alpha}^{-2})}\rho}\left(\frac{-3}{2\bar{\alpha}}+\frac{2}{
\alpha}+\frac{1}{re^{i\theta}-\alpha}+\frac{re^{i\theta}}{re^{i\theta}\bar{\alpha}-1}\right)+\mathcal{O}(\rho)
\end{align}
where the appropriate value is $\alpha=\sqrt{\frac{\rho}{2}}$:

\begin{align}
w(r,\theta)\sim \frac{re^{i\theta}(1+\frac{\rho}{2})+\sqrt{\frac{\rho}{2}}(1-3re^{2i\theta})}{40\sqrt{\frac{\rho}{2}}(1+\frac{\rho}{2})(\sqrt{\frac{\rho}{2}}-re^{i\theta})(\sqrt{\frac{\rho}{2}}re^{i\theta}-1)}
\end{align}

\section{Holography}
Having summarizes the three conformal maps under study, we move to discussing the holographic duals of the image of our setup under these maps. Our original worldsheet is a vacuum state so we expect its holographic dual to be empty Euclidean AdS$_3$ with the Poincar\'{e} metric:
\be 
\label{poincare}ds^2=\frac{dwd\bar{w}+dz^2}{z^2}
\ee
Where we have set the AdS radius to $R=1$ as is standard and the CFT lives on the boundary of AdS space at $z\rightarrow0$.

It is clear how the domain of the CFT changes under conformal mappings, but what happens to the bulk space under a conformal map? Ba\~nados \cite{Banados_1999} showed that the most general solution to Einstein's equations with negative cosmological constant (normalized to $-1$) in 2+1 dimensions is the following metric,
\be
ds^2=\frac{d\zeta^2}{\zeta^2}+L(\eta)d\eta^2+\bar{L}(\bar{\eta})d\bar{\eta}^2+\left(\frac{2}{\zeta^2}+\frac{\zeta^{2}}{2}L(\eta)\bar{L}(\bar{\eta})\right)d\eta d\bar{\eta},
\ee
where $\eta=\Xi+i\Upsilon,\bar{\eta}=\Xi-i\Upsilon$ are complex null coordinates of the theory living on the boundary of AdS, i.e. $\zeta\rightarrow0$, and $L,\bar{L}$ are arbitrary functions of $\eta$,$\bar{\eta}$.
The solution space is clearly infinite as it depends on the choice of two functions, both $L$ and $\bar{L}$.  Balasubramanian, et al. \cite{Balasubramanian_1999} later showed that a good definition for the holomorphic and anti-holomorphic stress tensors of asymptotically local AdS spacetimes is proportional to this choice of function: 
\be 
T(\eta)=\frac{-1}{16\pi }L(\eta), \bar{T}(\bar{\eta})=\frac{-1}{16\pi}\bar{L}(\bar{\eta}).
\ee
They further demonstrated that the stress tensor transforms as expected under conformal transformations:
\be
T(\eta)\rightarrow f'(\eta)T(f(\eta))+\frac{c}{12}\{f(\eta),\eta\}
\ee
where $c$ is the central charge of the CFT. Due to this, it is useful to consider a given stress tensor as the Schwarzian derivative of the conformal map from a vacuum solution since this implies $L(\eta)\propto \{f(\eta),\eta\}$ and $\bar{L}(\bar{\eta})\propto\{f(\bar{\eta}),\bar{\eta}\}$. 

It is often more convenient to work in the Poincar\'e patch of AdS$_3$. The appropriate coordinate transform from the Poincar\'e patch into the AdS$_3$ general solution was found by Roberts in 2012 \cite{Roberts_2012}. If we write the typical Poincar\'e metric as earlier Eq. (\ref{poincare}) the appropriate coordinate transform is:
\ba 
w &=& f(\eta)-\frac{2\zeta^2f'^2\bar{f}''}{4|f'|^2+\zeta^2|f''|^2}
\nonumber\\
\bar{w} &=& \bar{f}(\bar{\eta})-\frac{2\zeta^2\bar{f}'^2f''}{4|f'|^2+\zeta^2|f''|^2}
\nonumber\\
z &=& \frac{4\zeta (f'\bar{f}')^{3/2}}{4|f'|^2+\zeta^2|f''|^2}
\label{eq:bulk_transform}
\ea
So we can map our original world sheet for the splitting quench (left panel of \autoref{fig:3reps}) in coordinates $w,\bar{w},z$ to another setup with coordinates $\eta,\bar{\eta},\zeta$.

\subsection{Theta Function}\label{sec:theta_function}

In this section we reproduce the results of Caputa et al. \cite{caputa_double_2019} in which they use the modified theta function conformal map Eq. (\ref{eq:inverse_map_equation}), which is a map from the Jacobian variety to the doubly cut plane.
For this modified theta function we have:
\[\lim_{a\rightarrow 0}L(w(\eta))=-\pi^2\]
which gives
\be 
ds^2=\frac{R^2}{\zeta^2}\lb2\lb1-\frac{\pi^2\zeta^2}{2}\rb^2d\Xi^2+2
\lb1+\frac{\pi^2\zeta^2}{2}\rb^2d\Upsilon^2+d\zeta^2\rb\label{eq:BTZ_metric}.
\ee
Where $\Xi$ and $\Upsilon$ are periodic as shown in the middle section of \autoref{fig:3reps}, Eq. (\ref{eq:BTZ_metric}) is the BTZ metric\footnote{See Appendix A if this is not the form of the BTZ metric you are familiar with.}, and the event horizon occurs at $\zeta=\frac{\sqrt{2}}{\pi}$.

 We now proceed with the calculation of HEE of a spatial subsystem in the double splitting setup which is given by the minimal codimension two surface in the holographic dual which shares the same boundary as the subsystem \cite{ryu_holographic_2006}. Since the Euclidean holographic dual is conformally equivalent to the BTZ geometry, the computation of HEE consists in finding the geodesic length connecting the torus coordinates corresponding to the endpoints of the subsystem. A detailed derivation of the geodesic length connecting two points close to the BTZ boundary is performed in \autoref{sec:appendix_geodesics} and yields
\begin{align}
    L_\gamma
    &=\frac{1}{2}\ln\lb\lb\frac{2}{\epsilon_1\epsilon_2\pi^2}\rb^2\sin^2\lb\pi\Delta\eta_-\rb\sin^2\lb\pi\Delta\eta_+\rb\rb
\end{align}
where $\eta_-$ and $\eta_+$ are the analytical continuations of $\eta$ and $\Bar{\eta}$ respectively to recover the real time evolution of the system and $\Delta\eta_-=\eta_{-,1}-\eta_{-,2}$ where $1,2$ refer to the endpoints of the spatial subsystem $x_1,x_2$. $\epsilon_{1/2}$ are the bulk coordinate regulators corresponding to the torus coordinates $\eta_{1/2}$, respectively, and can be related to the UV cutoff $\epsilon$ in the original worldsheet via Eq. (\ref{eq:bulk_transform}):
\begin{align}
    \epsilon_{1/2}=\sqrt{2}\left|\frac{d\eta_{1/2}}{dw_{1/2}}\right|\epsilon
\end{align}
Here $w$ describes the complex coordinate on the original worldsheet.
Altogether this leads to the final result for the connected HEE, i.e.,
\begin{align}
    S_A^{\rm con}=\frac{c}{12}\ln\left[\lb\frac{1}{\epsilon\pi}\rb^4\frac{dw_{+,1}}{d\eta_{+,1}}\frac{dw_{-,1}}{d\eta_{-,1}}\frac{dw_{+,2}}{d\eta_{+,2}}\frac{dw_{-,2}}{d\eta_{-,2}}\sin^2\lb\pi\Delta\eta_-\rb\sin^2\lb\pi\Delta\eta_+\rb\right].\label{BTZ_Scon_final}
\end{align}

\begin{figure}[t]
         \centering
         \begin{subfigure}{0.45\textwidth}
            \includegraphics[width=\textwidth]{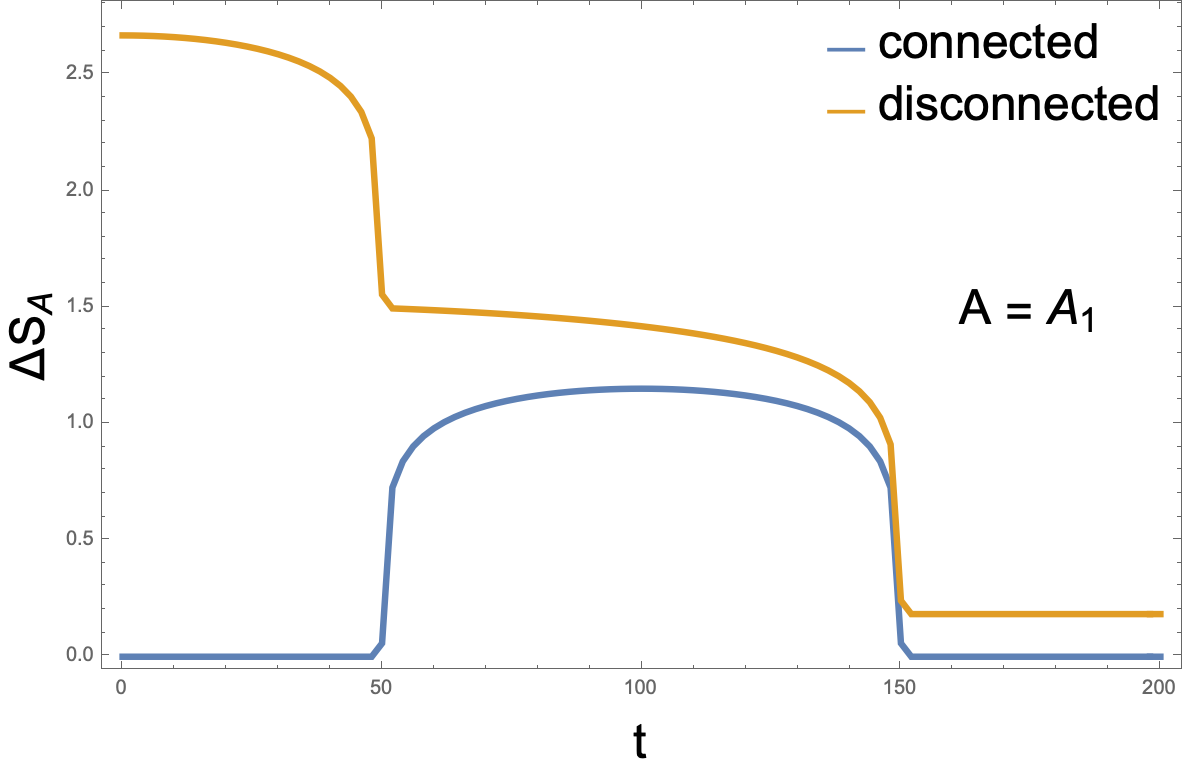}
         \end{subfigure}
         \hspace{0.05\textwidth}
         \begin{subfigure}{0.45\textwidth}
            \includegraphics[width=\textwidth]{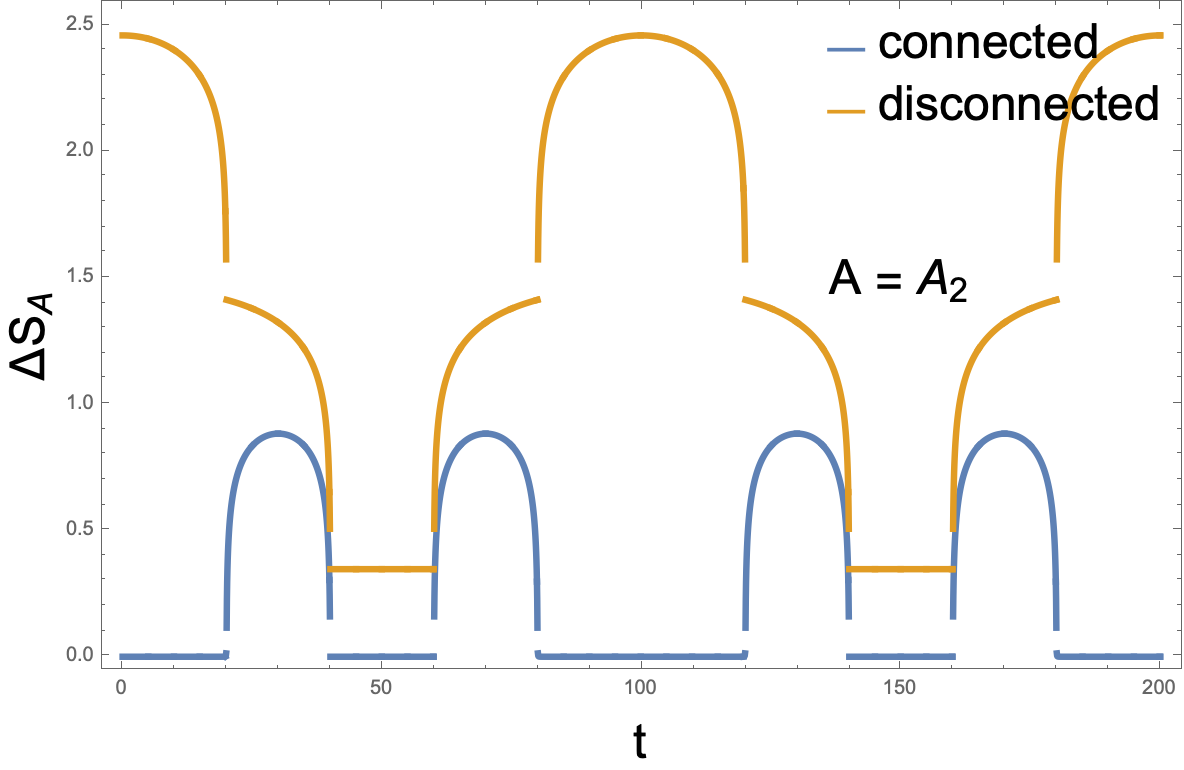}
         \end{subfigure}
         \begin{subfigure}{0.45\textwidth}
            \includegraphics[width=\textwidth]{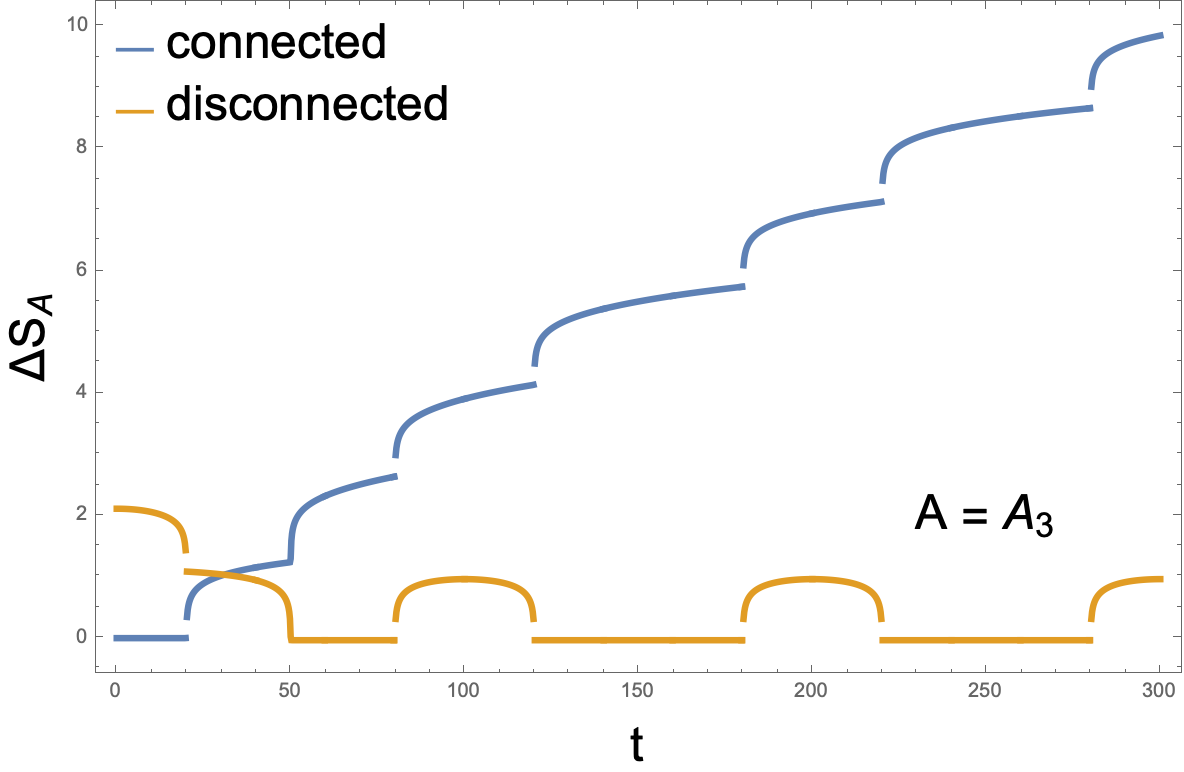}
         \end{subfigure}
         \hspace{0.05\textwidth}
         \begin{subfigure}{0.45\textwidth}
            \includegraphics[width=\textwidth]{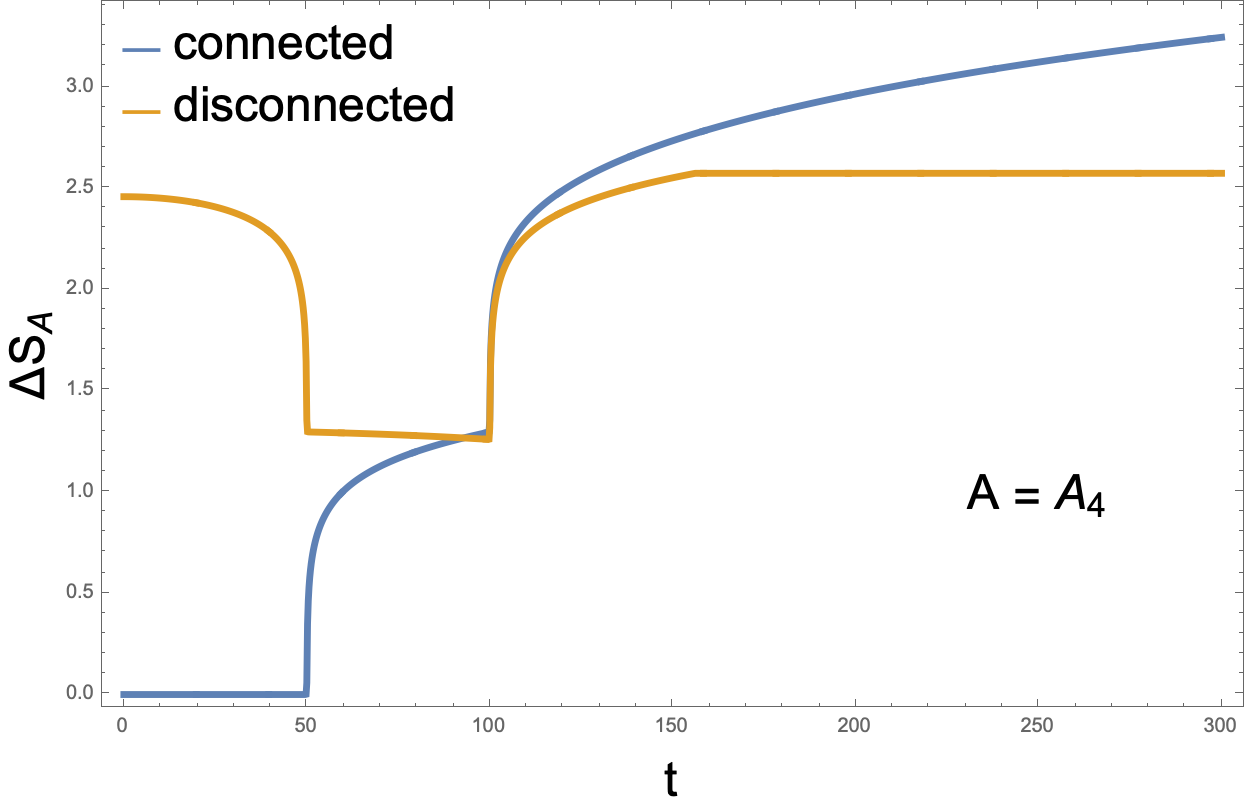}
         \end{subfigure}
         \caption{Numerical calculation of the holographic entanglement entropy growth $\Delta S_A$ for the subsystems $A_1$, $A_2$, $A_3$ and $A_4$ with parameters $b=50$, $a=0.05$ and $S_{{\rm bdy},1}=S_{{\rm bdy},2}=0$ using the theta function. The blue and orange lines are the connected and disconnected entropy growth respectively.}
         \label{fig:S_inv}
     \end{figure}

Apart from the connected geodesic there exists another possibility for a minimal path which does not violate the homology constraint, i.e. a disconnected geodesic.
In this case the path goes from one endpoint of the subsystem to one of the boundary surfaces located at $\Im(\eta)=\pm \tau/4$ and from the boundary to the second endpoint. Both parts of the geodesic can by disconnected and do not even have to be on the same boundary. In this case we have to introduce a horizon surface, connecting the two boundaries. This produces another term in the HEE as we will see below. Thus in order to calculate disconnected HEE we have to add the two minimal geodesics from both endpoints to some point of the boundary. The geodesic path lengths from some point on the BTZ conformal boundary to the BCFT boundaries are then given by one half of the geodesic lengths to the corresponding mirror points. The distance between some point and its mirror image is given by
\begin{align}
    L_{\pm}=\eta_--\eta_+\pm\frac{i
    \tau}{2}
\end{align}
where $\pm$ indicates the mirror point w.r.t the BCFT boundary at $\pm\frac{i\tau}{4}$.
The individual contributions to the HEE are again determined by the geodesic length derived in \autoref{sec:appendix_geodesics}. The final disconnected HEE is then given by the combination of smallest total entropy, i.e.
\begin{align}
    \begin{split}
        S_A^{\rm dis}=&\minl_{\sigma_1, \sigma_2=\pm}\left[\frac{c}{12}\ln\lb\frac{2}{\epsilon_1^2\pi^2}\sin^2\lb\pi L_{\sigma_1,1}\rb\rb+(1\leftrightarrow2)+\frac{1}{2}(1-\sigma_1\sigma_2)S_{\rm BH}\right]\\
        &+S_{{\rm bdy},1}+S_{{\rm bdy},2}\label{BTZ_disconnectedHEE}
    \end{split}
\end{align}
where $\frac{1}{2}(1-\sigma_1\sigma_2)$ equals $1$ when the two geodesic parts connect to different BCFT boundary surfaces. In that case an additional horizon must be added which yields in return the BTZ black hole entropy term
\begin{align}
    S_{\rm BH}=\frac{\rm Horizon\;Length}{4G}=\frac{\pi \tau}{4G}=\frac{c}{6}\pi \tau.\label{eq:BH_entropy}
\end{align}
where we used the identification $\frac{1}{4G}=\frac{c}{6}$ between the CFT and its gravity dual. In the case where both geodesic parts connect to the same boundary such a surface is not necessary and the black hole term is not added.
Finally, the holographic entanglement entropy of subsystem $A$ is given as the minimum of the connected and disconnected HEE. We show numerical calculations of the entanglement entropy growth $\Delta S_A=S_A-S_A^{(0)}$ in \autoref{fig:S_inv}, where $S_A^{(0)}=\frac{c}{3}\ln\frac{l}{\epsilon}$ is the well known vacuum entropy for a subsystem of length $l$ in 2d CFT. These results agree with those first presented in \cite{caputa_double_2019}.

\subsection{Abel-Jacobi map}

As the Abel-Jacobi Map is the inverse of the theta function, we find the same holographic entanglement entropy expressions as in \autoref{sec:theta_function} with the derivatives inverted:
\begin{align}
    S_A^{\rm con}=\frac{c}{12}\ln\left[\lb\frac{1}{\epsilon\pi}\rb^4\lb\frac{d\eta_{+,1}}{dw_{+,1}}\frac{d\eta_{-,1}}{dw_{-,1}}\rb^{-1}\lb\frac{d\eta_{+,2}}{dw_{+,2}}\frac{d\eta_{-,2}}{dw_{-,2}}\rb^{-1}\sin^2\lb\pi\Delta\eta_-\rb\sin^2\lb\pi\Delta\eta_+\rb\right]\label{BTZ_Scon_final_AJ}
\end{align}

\begin{figure}[t]
         \centering
         \begin{subfigure}{0.45\textwidth}
            \includegraphics[width=\textwidth]{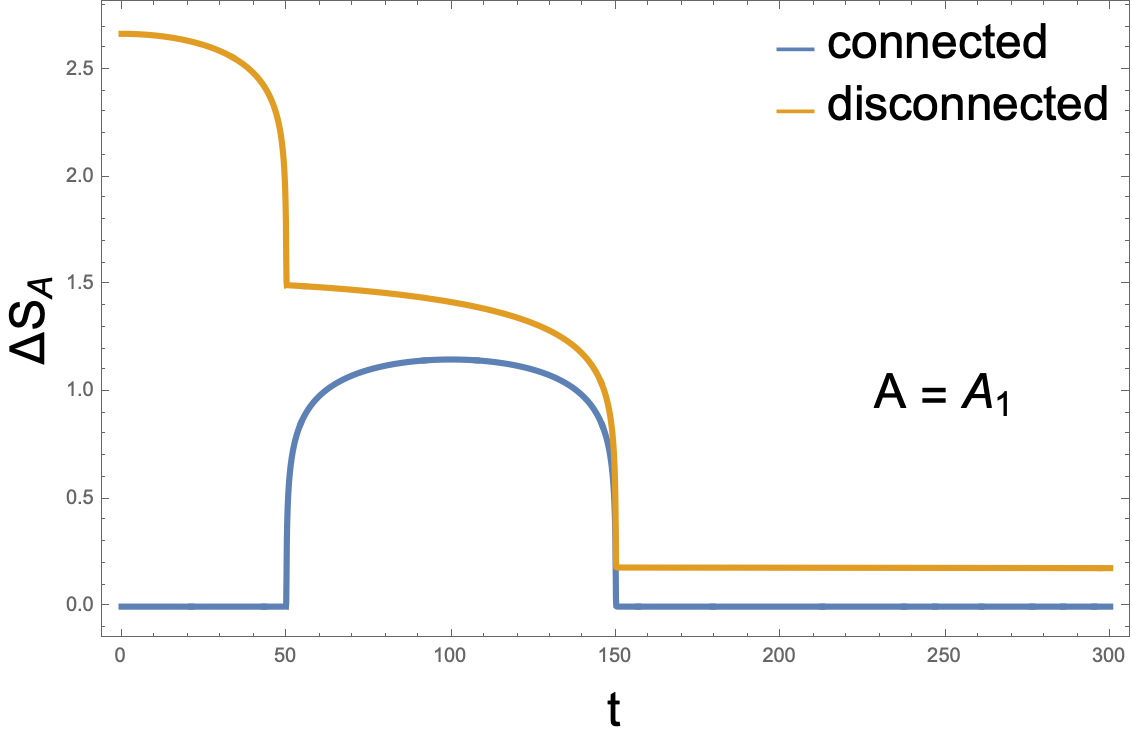}
         \end{subfigure}
         \hspace{0.05\textwidth}
         \begin{subfigure}{0.45\textwidth}
            \includegraphics[width=\textwidth]{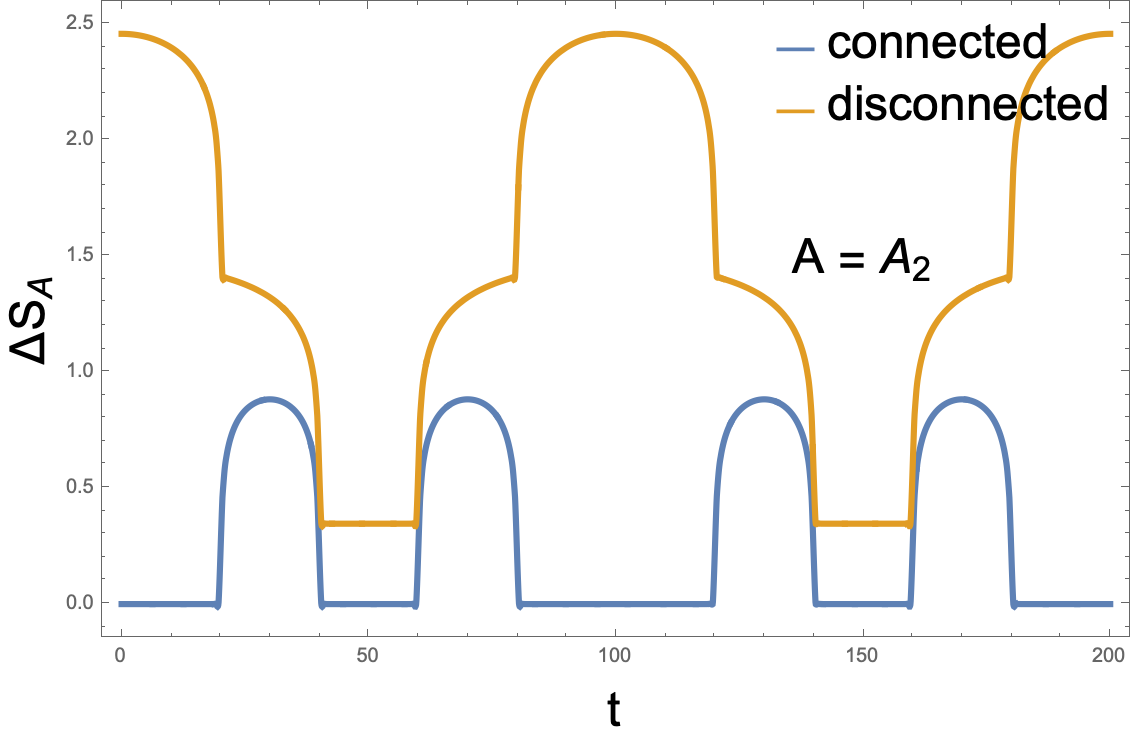}
         \end{subfigure}
         \begin{subfigure}{0.45\textwidth}
            \includegraphics[width=\textwidth]{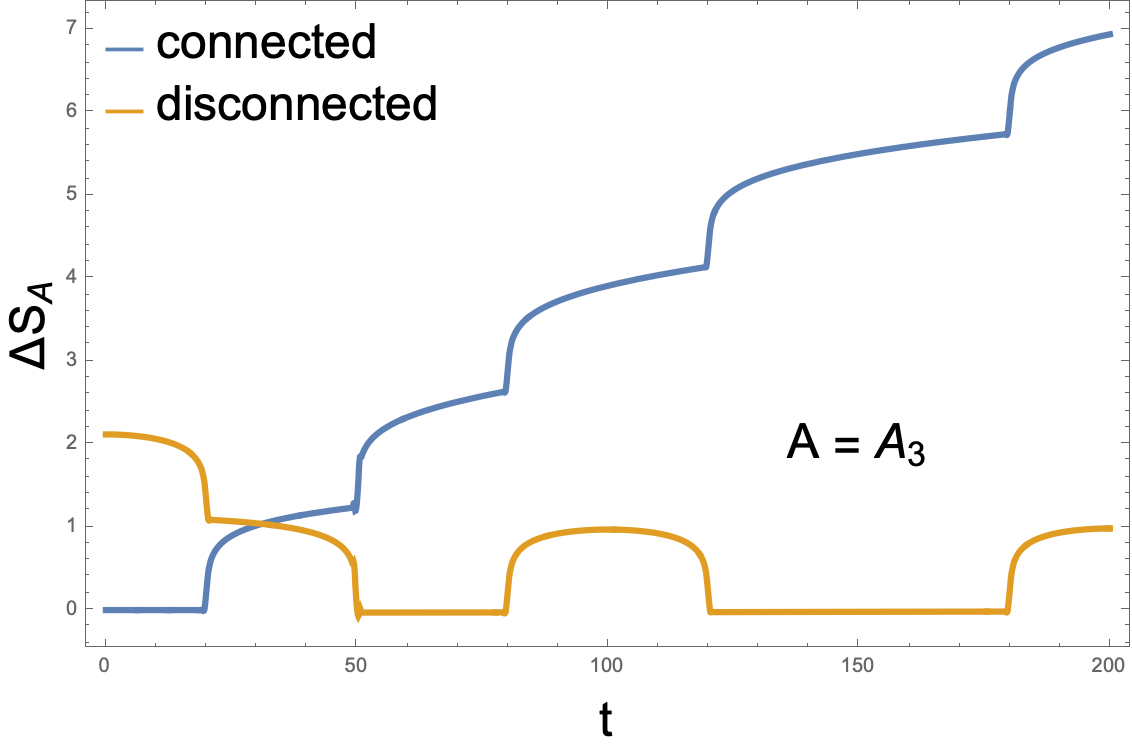}
         \end{subfigure}
         \hspace{0.05\textwidth}
         \begin{subfigure}{0.45\textwidth}
            \includegraphics[width=\textwidth]{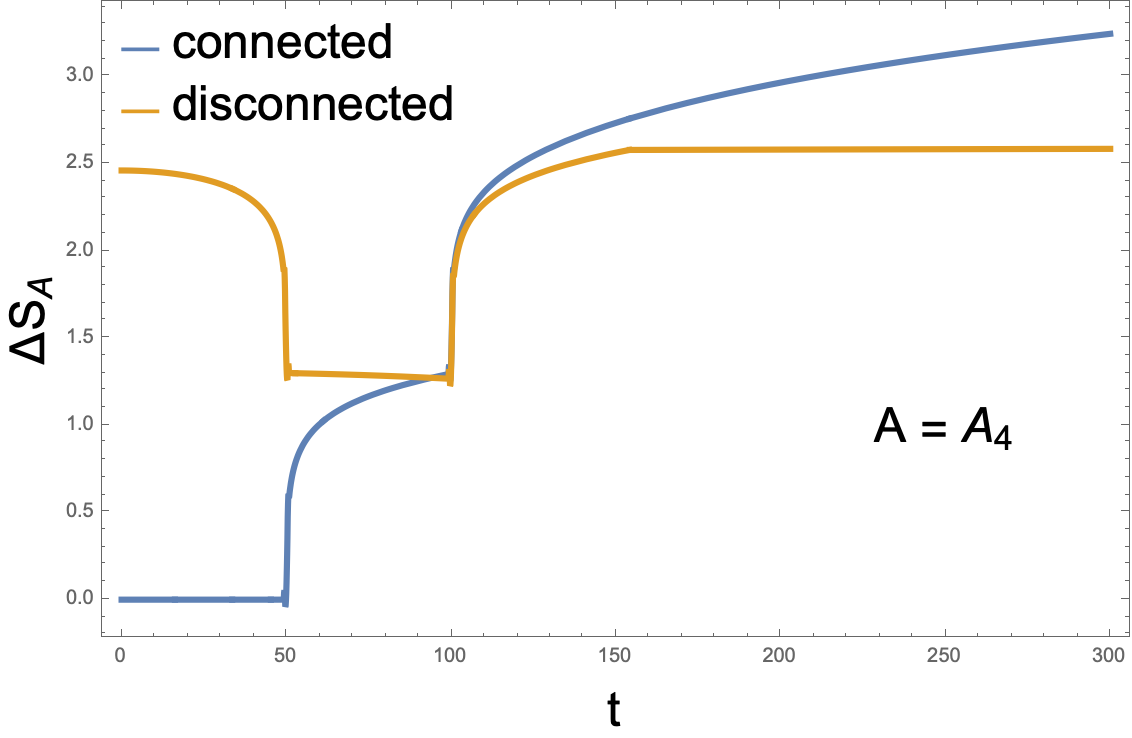}
         \end{subfigure}
         \caption{Numerical calculation of the holographic entanglement entropy growth $\Delta S_A$ for the subsystems $A_1$, $A_2$, $A_3$ and $A_4$ with parameters $b=50$, $a=0.05$ and $S_{{\rm bdy},1}=S_{{\rm bdy},2}=0$ using the Abel-Jacobi map. The blue and orange lines are the connected and disconnected entropy growth respectively.}
         \label{fig:S_AJ}
    \end{figure}

Similarly we obtain the disconnected HEE. In \autoref{fig:S_AJ} we depict the connected and disconnected holographic entanglement entropy computed for for the various subsystem setups using the Abel-Jacobi method. We find perfect agreement for all analyzed subsystems, which demonstrates the equivalence between the inverse map method and the Abel-Jacobi procedure.

When generalizing the map Eq. (\ref{eq:inverse_map_equation}) to a larger number of cuts, one would have to deal with generalized theta functions, which are famously cumbersome. The set of canonical differentials and Abel-Jacobi maps from this approach are just as easily defined on hyperelliptic curves as they are on elliptic curves.

\subsection{Schottky uniformization}

\begin{figure}
    \centering
    \tdplotsetmaincoords{60}{110}

\pgfmathsetmacro{\radius}{1}
\pgfmathsetmacro{\thetavec}{0}
\pgfmathsetmacro{\phivec}{0}
\pgfmathsetmacro{\r}{1/3}
\pgfmathsetmacro{\rcoord}{0.15}

\begin{tikzpicture}[scale=4.5,tdplot_main_coords]

\draw [orange, ultra thick] (0,-\r,0) node [right] {\color{black}$\rho$} --(0,-3*\r,0)node [left] {\color{black}$\frac{1}{\rho}$} node [pos=0.5,above right] {\color{black}$1$};
\draw [yellow, ultra thick] (0,\r,0)--(0,1.5*\r,0);
\draw [blue, ultra thick] (0,1.5*\r,0)--(0,2.5*\r,0);
\draw [yellow, ultra thick] (0,2.5*\r,0)--(0,3*\r,0);
\node [cross, label=above:{$\alpha$}] (a) at (0,1.5*\r,0) {};
\node [cross, label=above:{$\frac{1}{\alpha}$}] (a) at (0,2.5*\r,0) {};

\tdplotsetthetaplanecoords{\phivec}

\fill [cyan!50!gray, semitransparent,even odd rule] (0,0,0) circle[radius=\r] circle[radius=2*\r];
\fill [blue!50!gray, fill opacity=0.5,even odd rule] (0,0,0) circle[radius=2*\r] circle[radius=3*\r];

\draw[thick,->] (0,0,0) -- (0,\rcoord,0) node[below]{$r$};
\draw[thick,->] (0,0,0) -- (0,0,\rcoord) node[anchor=south]{$\xi$};
\draw[thick,->] (90:\rcoord) arc (90:360:\rcoord) node[below]{$\theta$};
\shade[ball color=red,opacity=0.25] (0.333cm,0) arc (0:-180:0.333cm and 1.667mm) arc (180:0:0.333cm and 0.333cm);
\draw[red,thick,<->] (0,0,\r) -- (0,0,1) node[pos=.5,right]{$\text{identified}$};
\shade[ball color=red,opacity=0.25] (1cm,0) arc (0:-180:1cm and 5mm) arc (180:0:1cm and 1cm);
\end{tikzpicture}
    \caption{The Schottky uniformization of the genus 1 surface. The inner and outer red hemispheres are identified forming the BTZ blackhole with the annulus as its boundary.}
    \label{fig:SchottkyHolo}
\end{figure}
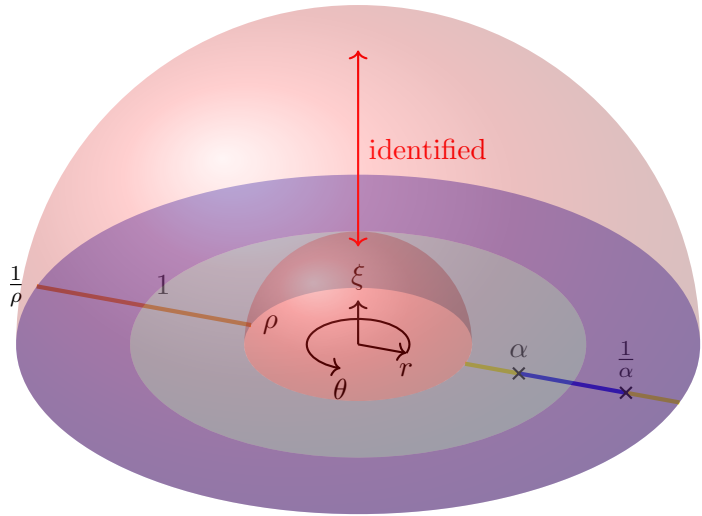

We use a slight variation of a procedure developed in \cite{Krasnov2000} to find the holographic dual of compact Riemann surfaces. Consider a genus $g$ hyperelliptic curve, it can be uniformized by a Schottky domain, where one sheet is a collection of $g$ circles cut out of the unit disk, and the Schottky double (other sheet) is the reflection of the interior of the unit disk across the unit circle. For the case of $g=1$ we have an annulus of inner radius $\rho<1$ (which is a function of $a$ and $b$) and unit outer radius. When doubled the outer radius $1/\rho$ is identified with the inner radius (See \autoref{fig:SchottkyHolo}). The identification of the inner to outer circle is a M\"{o}bius transform and therefore the group it generates is a subgroup of $SL(2,\mathbb{C})$. We can then quotient $\mathbb{H}^3$ space by this group to obtain our geometry, which is locally Euclidean AdS$_3$ with the genus $g$ surface as its boundary. For the case of $g=1$ this is the familiar BTZ black hole. While there is the standard BTZ to thermal AdS transition as shown in \cite{caputa_double_2019}, as we are only considering the $a\rightarrow0$ limit we are in the region of moduli space where we only need to consider BTZ.

The Schwarzian derivative of the map from the annulus to the plane with two cuts is not globally defined due to our mirroring procedure. We make use of the fact that the map from the identified rectangle to the annulus is $\sqrt{\rho}e^{2\pi i \eta}$ (with $\rho=e^{-\pi \tau}$). Then the Schwarzian derivative of the composite map is globally defined:
\[
L(w(r(\eta),\theta(\eta)))=-\pi^2
\]
which leads to the metric:
\[
\frac{ds^2}{R^2}=-\pi^2d\eta^2-\pi^2d\bar{\eta}^2+\frac{4+z^4\pi^4}{2z^2}d\eta d\bar{\eta}+\frac{dz^2}{z^2}.
\]
Expressing  this in $\theta$ and $\phi=\ln(r)$ yields:
\[ds^2=\frac{R^2}{z^2}\lb\frac{(\frac{\pi^2}{2}z^2-1)^2}{2\pi^2}d\theta^2+\frac{(\frac{\pi^2}{2}z^2+1)^2}{2\pi^2}d\phi^2+dz^2\rb\]
Which we note is the same BTZ metric as Eq. (\ref{eq:BTZ_metric}) if we scale both angular coordinates by $2\pi$.
The geodesic is then the same as the one calculated in Appendix B, and the entanglement entropy is the same as Eq. (\ref{BTZ_Scon_final}).

\section{Conclusion}

We have introduced a novel formalism for holographic duals of non-simply connected world sheets. The formalism is to conformally map from the Schottky domain of the world sheet to the worldsheet using maps derived from the Schottky-Klein prime function \cite{crowdy2020solving}. The extension of the conformal map into the bulk coordinate is computed via the standard prescription 
\cite{Banados_1999,Balasubramanian_1999,Roberts_2012} and we follow \cite{Krasnov2000,Krasnov2002,Krasnov2003} in identifying the extensions into the bulk of the circles. This allows us to calculate entanglement entropy as a function of time (first Euclidean time and after continuation Lorentzian time) and other dynamical observables with the holographic entanglement entropy formalism.

In this paper we have shown that this formalism applied to the case of a double splitting quench reproduces earlier results \cite{caputa_double_2019}. The use of the Abel-Jacobi map greatly simplifies the calculations. In a forthcoming paper we will present explicit calculations of the holographic entanglement entropy for quenches with multiple splits and we hope to apply the formalism to other instances of non-simply connected worldsheets, such as splitting quenches at finite temperature and multiple local projection measurements.
In addition, we hope to use these ingredients to build a schematic model of the multifragmentation process in heavy ion collisions: Analyzing the entanglement structure among multiple ``hadrons'' that split off from a highly excited, locally thermal pure state of a ``quark-gluon plasma.''

\acknowledgments
We would like to thank Bertrand Eynard for discussions on Riemann surfaces and Zixia Wei for discussions on splitting quenches. JDL and BM were supported by a research grant (DE-FG02-05ER41367) from the U.S. Department of Energy Office of Science. BM also acknowledges support from Yale University during extended visits in 2022 and 2023. CS acknowledges support from Duke University during a visit in 2023.

\appendix

\section{Different Coordinates for the BTZ Geometry}
\label{sec:appendix_BTZ}

We claim the metric in Eq. (\ref{eq:BTZ_metric}) is the BTZ metric. Here we show it's equivalent to the more standard representation.
\[ds^2=\frac{R^2}{z^2}\lb\lb1-\frac{\pi^2z^2}{2}\rb^2dx^2+\lb1+\frac{\pi^2z^2}{2}\rb^2dy^2+dz^2\rb\]
Where $x\in(0,1],y\in (-\tau/2,\tau/2)$ are both periodic and $z\in[0,\infty)$\\
Consider the coordinate transform:
\[z'=\frac{z}{(1+\frac{\pi^2z^2}{2})}\]
Where $z' \in [0,\frac{1}{\sqrt{2}\pi}]$.\\
Note that the conformal boundary is left invariant.\\
The metric then becomes:
\[ds^2=\frac{R^2}{z'^2}\lb\lb1-2\pi^2z'^2\rb dx^2+dy^2+\frac{dz'^2}{1-2\pi^2z'^2}\rb\]

Which is the form of the metric we use for our geodesic calculations.

Consider an additional coordinate transform $z'=\frac{1}{\sqrt{z''^2+2\pi^2}}$\\

\[ds^2=R^2 (z''^2+2\pi^2)\lb\frac{z''^2}{z''^2+2\pi^2}dx^2+dy^2+\lb\frac{z''^2+2\pi^2}{z''^2}\rb\frac{z''^2dz''^2}{(z''^2+2\pi^2)^3}\rb\]

\[ds^2=R^2\lb z''^2dx^2+(z''^2+2\pi^2)dy^2+\frac{dz''^2}{z''^2+2\pi^2}\rb\]
This looks odd but replacing the periodic coordinates with angles, the radial coordinate with r, and defining $r_s^2=2\pi^2$ gives:
\[ds^2=R^2\lb(r^2+r_s^2)dt^2+\frac{dr^2}{(r^2+r_s^2)}+r^2d\phi^2\rb\]
Which is the most standard presentation of the Euclidean BTZ metric.

\section{Calculation of the Geodesic length}
\label{sec:appendix_geodesics}

In this section we calculate the geodesic length between two points in BTZ space close to the boundary $z\rightarrow0$.
As explained in \autoref{sec:appendix_BTZ} the BTZ metric Eq. (\ref{eq:BTZ_metric}) can be brought into the form
\begin{align}
    ds^2=\frac{1}{z^2}\left[\frac{dz^2}{1-\alpha^2z^2}+(1-\alpha^2z^2)\;dx^2+dy^2\right]
\end{align}
by an appropriate coordinate transformation, where $\alpha=\sqrt{2}\pi$.
The length of some path $\gamma$ is thus given by
\begin{align}
    L_\gamma=\intl_\gamma dz\underbrace{\frac{1}{z}\sqrt{\frac{1}{1-\alpha^2z^2}+(1-\alpha^2z^2)(x')^2+(y')^2}}_{\mathcal{L}(z,x',y')}
\end{align}
Since the geodesic path $\gamma_A$ is a stationary point of $L_\gamma$, the Euler-Lagrange equations have to be fulfilled along $\gamma_A$ where $\mathcal{L}(z,x',y')$ is the corresponding Lagrangian. Using the fact that the Lagrangian is independent of $x$ and $y$, we obtain
\begin{align}
    c_1=\frac{\partial\mathcal{L}}{\partial x'}, \hspace{1cm}c_2=\frac{\partial\mathcal{L}}{\partial y'}.
\end{align}
Solving these equations for $T$ and $X$ yields
\begin{align}
    \begin{aligned}
        x'(z)&=\frac{c_1z}{(1-\alpha^2z^2)\sqrt{1-(c_1^2+c_2^2+\alpha^2)z^2+c_2^2\alpha^2z^4}}\\\
        y'(z)&=\frac{c_2z}{\sqrt{1-(c_1^2+c_2^2+\alpha^2)z^2+c_2^2\alpha^2z^4}}
    \end{aligned}
\end{align}
Because of the symmetry of BTZ space, we assume the geodesic to be symmetric about some maximum value $z_*$ for endpoints on the boundary. At this point, the partial derivatives of $z(x,y)$ w.r.t $x$ and $y$ are zero by definition which means in return that $x'(z_*)\rightarrow\infty$ and $y'(z_*)\rightarrow\infty$.
From this we find
\begin{align}
    c_1=\frac{1}{z_*}\sqrt{1-c_2^2z_*^2}\sqrt{1-\alpha^2z_*^2}
\end{align}
and the derivatives take the form
\begin{align}
    \begin{aligned}
        x'(z)&=\lb\frac{1}{z_*}\sqrt{1-c_2^2z_*^2}\sqrt{1-\alpha^2z_*^2}\rb\frac{z}{(1-\alpha^2z^2)\sqrt{(1-\frac{z^2}{z_*^2})(1-c_2^2\alpha^2z_*^2z^2)}}\\\
        y'(z)&=c_2\frac{z}{(1-\alpha^2z^2)\sqrt{(1-\frac{z^2}{z_*^2})(1-c_2^2\alpha^2z_*^2z^2)}}
    \end{aligned}
\end{align}
We now calculate the differences $\Delta T$ and $\Delta X$ of the endpoints on the boundary. Since we assume the geodesic to be symmetric about $z_*$, the path from $z=\epsilon$ to $z=\epsilon$ is just given by twice the integral from $z=\epsilon$ to $z=z_*$.
\begin{align}
    \begin{split}
        \Delta x
        &=2\intl_\epsilon^{z_*}dz\;x'(z)=\frac{2}{\alpha}\arctanh\lb\frac{\alpha z_*\sqrt{1-c_2^2z_*^2}}{\sqrt{1-\alpha^2z_*^2}}\rb\label{eq:delta_x}
    \end{split}\\
    \begin{split}
        \Delta y
        &=2\intl_\epsilon^{z_*}dz\;y'(z)=\frac{2}{\alpha}\arctanh(c_2\alpha z_*^2)\label{eq:delta_y}
    \end{split}
\end{align}
From Eq. (\ref{eq:delta_y}) follows
\begin{align}
    c_2=\frac{\tanh(\frac{\alpha\Delta y}{2})}{\alpha z_*^2}
\end{align}
which we plug into Eq. (\ref{eq:delta_x}) to solve for the maximum value
\begin{align}
    z_*=\frac{\sqrt{\tanh^2\left(\frac{\alpha\Delta y}{2}\right)+\tan^2\left(\frac{\alpha\Delta x}{2}\right)}}{\alpha\sqrt{1+\tan^2\left(\frac{\alpha\Delta x}{2}\right)}}
\end{align}

Thus we have determined all constants $c_1$, $c_2$, $z_*$ and we can now calculate the geodesic length, where we define $\beta:=c_2\alpha z_*$:
\begin{align}\notag
        L_\gamma
        &=\intl_\gamma dz\frac{1}{z}\sqrt{\frac{1}{1-\alpha^2z^2}+(1-\alpha^2z^2)(x')^2+(y')^2}\\
        &=\frac{1}{\beta}\lb\intl_{\epsilon_1}^{z_*}dz\frac{1}{z\sqrt{(\frac{z^2}{z_*^2}-1)(z^2-\frac{1}{\beta^2})}}+\intl_{\epsilon_2}^{z_*}dz\frac{1}{z\sqrt{(\frac{z^2}{z_*^2}-1)(z^2-\frac{1}{\beta^2})}}\rb\notag\\\
        \notag
        &=\ln\left(\frac{-\frac{2z_*}{\beta}\sqrt{(z^2-z_*^2)(z^2-\frac{1}{\beta^2})}-\frac{2z_*^2}{\beta^2}+z_*^2z^2+\frac{z^2}{\beta^2}}{z^2(z_*^2-\frac{1}{\beta^2})}\right)\Bigg|_{z=\epsilon_1}^{z=z_*}
        +(\epsilon_1\leftrightarrow\epsilon_2)\\\
        \notag
        &=-\ln(\epsilon_1)-\ln(\epsilon_2)+\ln\lb\frac{4z_*^2}{1-\beta^2z_*^2}\rb+\mathcal{O}\lb\epsilon^2\rb\\\
        &=\frac{1}{2}\ln\lb\lb\frac{4}{\epsilon_1\epsilon_2\alpha^2}\rb^2\sin^2\lb\frac{\alpha}{2}(\Delta x-i\Delta y)\rb\sin^2\lb\frac{\alpha}{2}(\Delta x+i\Delta y)\rb\rb
    \end{align}

\bibliographystyle{JHEP}
\bibliography{main.bib}

\nocite{*}

\end{document}